\documentclass[12pt]{iopart}

\usepackage[separate-uncertainty=true, print-zero-exponent=false, print-unity-mantissa=false]{siunitx}
\DeclareSIUnit\angstrom{\text{Å}}
\usepackage[utf8]{inputenc}
\usepackage{xcolor}
\usepackage{graphicx}
\usepackage{float}
\usepackage{amsmath}
\usepackage{multirow}
\usepackage{amssymb}
\usepackage{tikz}
\usepackage[breaklinks=true, colorlinks]{hyperref}
\usepackage[noabbrev]{cleveref}
\usepackage[smartEllipses]{markdown}
\usepackage{soul}
\usepackage{booktabs}
\usepackage[numbers]{natbib}
\usepackage[normalem]{ulem}

\let\oldequation\equation
\let\oldendequation\endequation

\renewenvironment{equation}
  {\linenomathNonumbers\oldequation}
  {\oldendequation\endlinenomath}

\definecolor{tud0d}{RGB}{83,83,83}
\definecolor{tud0c}{RGB}{137,137,137}
\definecolor{tud0b}{RGB}{181,181,181}
\definecolor{tud0a}{RGB}{220,220,220}

\definecolor{tud1a}{RGB}{93,133,195}
\definecolor{tud2a}{RGB}{0,156,218}
\definecolor{tud3a}{RGB}{80,182,149}
\definecolor{tud4a}{RGB}{175,204,80}
\definecolor{tud5a}{RGB}{221,223,72}
\definecolor{tud6a}{RGB}{255,224,92}
\definecolor{tud7a}{RGB}{248,186,60}
\definecolor{tud8a}{RGB}{238,122,52}
\definecolor{tud9a}{RGB}{233,80,62}
\definecolor{tud10a}{RGB}{201,48,142}
\definecolor{tud11a}{RGB}{128,69,151}

\definecolor{tud1b}{RGB}{0,90,169}
\definecolor{tud2b}{RGB}{0,131,204}
\definecolor{tud3b}{RGB}{0,157,129}
\definecolor{tud4b}{RGB}{153,192,0}
\definecolor{tud5b}{RGB}{201,212,0}
\definecolor{tud6b}{RGB}{253,202,0}
\definecolor{tud7b}{RGB}{245,163,0}
\definecolor{tud8b}{RGB}{236,101,0}
\definecolor{tud9b}{RGB}{230,0,26}
\definecolor{tud10b}{RGB}{166,0,132}
\definecolor{tud11b}{RGB}{114,16,133}

\definecolor{tud1c}{RGB}{0,78,138}
\definecolor{tud2c}{RGB}{0,104,157}
\definecolor{tud3c}{RGB}{0,136,119}
\definecolor{tud4c}{RGB}{127,171,22}
\definecolor{tud5c}{RGB}{177,189,0}
\definecolor{tud6c}{RGB}{215,172,0}
\definecolor{tud7c}{RGB}{210,135,0}
\definecolor{tud8c}{RGB}{204,76,3}
\definecolor{tud9c}{RGB}{185,15,34}
\definecolor{tud10c}{RGB}{149,17,105}
\definecolor{tud11c}{RGB}{97,28,115}

\definecolor{tud1d}{RGB}{36,53,114}
\definecolor{tud2d}{RGB}{0,78,115}
\definecolor{tud3d}{RGB}{0,113,94}
\definecolor{tud4d}{RGB}{106,139,55}
\definecolor{tud5d}{RGB}{153,166,4}
\definecolor{tud6d}{RGB}{174,142,0}
\definecolor{tud7d}{RGB}{190,111,0}
\definecolor{tud8d}{RGB}{169,73,19}
\definecolor{tud9d}{RGB}{156,28,38}
\definecolor{tud10d}{RGB}{115,32,84}
\definecolor{tud11d}{RGB}{76,34,106}

\definecolor{tab10_0}{HTML}{1f77b4}
\definecolor{tab10_1}{HTML}{ff7f0e}
\definecolor{tab10_2}{HTML}{2ca02c}
\definecolor{tab10_3}{HTML}{d62728}
\definecolor{tab10_4}{HTML}{9467bd}
\definecolor{tab10_5}{HTML}{8c564b}
\definecolor{tab10_6}{HTML}{e377c2}
\definecolor{tab10_7}{HTML}{7f7f7f}
\definecolor{tab10_8}{HTML}{bcbd22}
\definecolor{tab10_9}{HTML}{17becf}

\newcommand{\sio}[0]{SiO\ensuremath{_2}}
\newcommand{\sioTitle}[0]{SiO\texorpdfstring{\ensuremath{_2}}{2}}

\usepackage{marginnote}

\begin{document}

\def\thefootnote{*}\footnotetext{These authors contributed equally to this work}\def\thefootnote{\arabic{footnote}}

\title[Crystal structure identification using 3D convolutional neural networks]{Crystal structure identification with 3D convolutional neural networks with application to high-pressure phase transitions in \sioTitle{}}
\
\author{Linus C.\ Erhard$^\dagger$$^1$}

\author{Daniel Utt$^\dagger$$^{2,1}$}

\author{Arne J.\ Klomp$^1$}

\author{Karsten Albe$^1$}
´
\address{$^1$Fachgebiet
Materialmodellierung, Institut f\"ur Materialwissenschaft, Technische Universit\"at Darmstadt, Otto-Berndt-Str.\ 3, D-64287 Darmstadt, Germany}
\address{$^2$SDL Materials Design, NHR4CES, Technische Universit\"at Darmstadt, D-64287 Darmstadt, Germany}
\ead{erhard@mm.tu-darmstadt.de}
\ead{albe@mm.tu-darmstadt.de}

\begin{abstract}

    Efficient, reliable and easy-to-use structure recognition of atomic environments is essential for the analysis of atomic scale computer simulations. 
    In this work, we train two  neuronal network (NN) architectures, namely PointNet and dynamic graph convolutional NN (DG-CNN) using different hyperparameters and training regimes to assess their performance in structure identification tasks of atomistic structure data. We show benchmarks on \textit{simple} crystal structures, where we can compare against established methods. The approach is subsequently extended to structurally more  complex \sio{} phases. By making use of this structure recognition tool, we are able to achieve a deeper understanding of the crystallization process in amorphous \sio{} under shock compression. 
    Lastly, we show how the NN based structure identification workflows can be integrated into \textsc{OVITO}  using its \textsc{python} interface.
    
\end{abstract}

\maketitle

\section{Introduction}

Atomistic simulations have become an invaluable tool in materials research, as they allow for the correlation of observations on the atomic scale with macroscopic properties.
To understand these observations it is essential to analyze the underlying local structures within the simulations.
This analysis should be able to detect distinct crystallographic phases and lattice defects, which cause changes in macroscopic properties. 
Common algorithms can identify  typical crystal structures, like face or body centered cubic (FCC / BCC), cubic diamond (cDia), or hexagonal closest packed (HCP) lattices \cite{Honeycutt1987,Faken1994,Ackland2006,Stukowski2012a,Larsen2016,Maras2016} or even crystalline H\(_2\)O (ice) \cite{Nguyen2015} based on the atomic positions.
These methods usually involve the calculation of an order parameter, which is then used to label atoms according to a specific crystal structure based on mathematical descriptors.

Recently, the process of finding such descriptors has been delegated more and more to machine learning (ML) methods in order to accelerate the development of new structure identification algorithms and improve their accuracy. 

One method evolves around batching of atomistic data into smaller discrete voxels. These smaller atomic groups are characterized using simulated diffraction patterns, which are subsequently processed by a image classification neural network (NN) \cite{Ziletti2018}. This approach was later on refined, replacing the simulated diffraction patterns with the smooth overlap of atomic orbitals (SOAP) descriptor \cite{Bartok2010, Bartok2013}. This descriptor is fed into a NN classifier to identify the different crystal structures \cite{Leitherer2021}. In both cases, the diffraction pattern and SOAP descriptors act as an abstraction method to compress the atomistic coordinates into a format that is more digestible by the NN classifiers. However, one downside is that crystal structure classification information is not a property assigned to each atom but instead to each voxel, leading to a coarse graining of the output. This is different to established analytical methods, like polyhedral template matching (PTM) \cite {Larsen2016} or common neighbor analysis (CNA),\cite{Honeycutt1987,Faken1994,Stukowski2012a}, which provide structural information for every single atom. 

An alternative approach was presented in a recent work by Chung et al.\cite{Chung2022}. They  first filter out non-crystalline atoms, calculate a feature vector for the remaining atoms and obtain a classification using a small feed-forward NN. This network predicts the local crystal structure for each atom and a final filtering step corrects for unknown crystal structures using again a heuristic criterion.
There exist similar approaches where the local atomic descriptors are combined with a ML algorithm to identify vacancies and other defects \cite{Goryaeva2020} or different types of amorphous structures \cite{beaulieu2024} in simulated samples. 

An alternative to these methods is not to rely on any descriptors and to directly use the atomic positions as an input into an ML method. Here one could directly apply the neural network architectures used for computer vision based on 3D point cloud data as is done for interpreting, e.g., LIDAR data. One such example is the work by DeFever et al.\cite{DeFever2019}. They use the PointNet NN\cite{Qi2016}, which is one of the NNs directly operating on 3D spatial point data, to classify atomic environments. This NN classifier was trained on molecular dynamics (MD) trajectory data and delivered very good classification accuracy. In Ref.~\citenum{DeFever2019} the authors decided to include all atoms within a given cutoff radius around the central atom in the local environment. The NN, however, needs a constant number of points as input. This means that this atomic positions vector has to be truncated or padded depending on the local density, which might influence the classification accuracy.

\begin{figure*}
    \centering
    \includegraphics{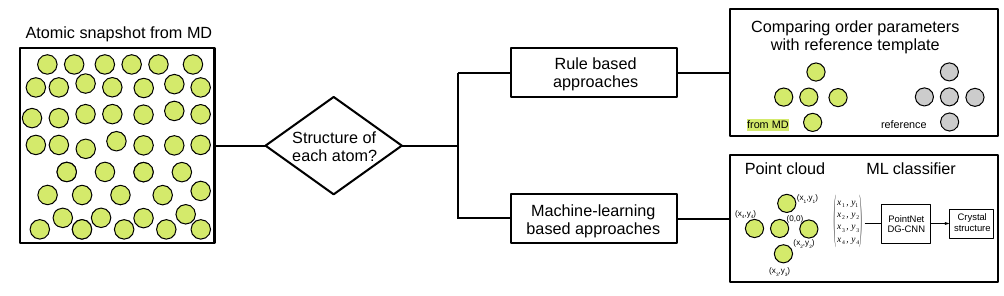} 
    \includegraphics{preprint_figures/Fig1_illustration.pdf} 
    \caption{\textbf{Comparison of classical and machine-learning structure identification.} The problem of determining the crystal structure of individual atoms within MD simulations can be solved by two different approaches nowadays. There is the classical way of using rule based approaches, like CNA or PTM. These approaches are implemented well in current codes and are fast to evaluate, however, are restricted to simple crystal structures, e.g.\ BCC, FCC, HCP. Defining additional rules for other structures is complicated and time-consuming. In contrast, machine-learning approaches can be easily trained to a large number of artificially generated training data with the corresponding crystal structures and can thus be extended easily to other more complex crystal structures.}
    \label{fig:intro}
\end{figure*}

An overview of the rule based approach compared to the last ML based approach is illustrated in Fig. \ref{fig:intro}. For simple crystal systems, there are already mature crystal structure identification schemes based on simple rules. However, there is a high demand for reliable and easy to use structure recognition algorithms for more complex system. An example for a structurally challenging system with high relevance for science is the \sio{} system. This material exhibits a high degree of polymorphism at ambient conditions as well under high pressures \cite{heaneySilicaPhysicalBehavior1994a}. Due to its importance for earth science many simulations and experiments have been performed, which proposed a significant amount of hypothetical high-pressure structures \cite{badroTheoreticalStudyFivecoordinated1997,choudhuryInitioStudiesPhonon2006,liuNewHighpressureModifications1978,tseHighpressureDensificationAmorphous1992,teterHighPressurePolymorphism1998,wentzcovitchNewPhasePressure1998,svishchevOrthorhombicQuartzlikePolymorph1997,otzenEvidenceRosiaitestructuredHighpressure2023}. However, up to now, there has been no framework capable of differentiating all these phases, making it difficult to recognize the structures in MD simulations, such as those of shock compressions.

NNs seem promising in solving the complex structure identification task in \sio{}. In this work, we train two  NN architectures, namely PointNet\cite{Qi2016} and dynamic graph convolutional NN (DG-CNN) \cite{Wang2019} using different hyperparameters and training regimes to assess their performance in structure identification tasks. We show benchmarks on \textit{simple} crystal structures, where we can compare against established methods. The approach is subsequently tested using more complex SiO$_2$ structures as an example. Lastly, we show how  NN based structure identification workflows can be integrated into \textsc{ovito} \cite{Stukowski2010} using its \textsc{python} interface.

\section{Methodology}

\begin{figure}
    \centering
    \includegraphics{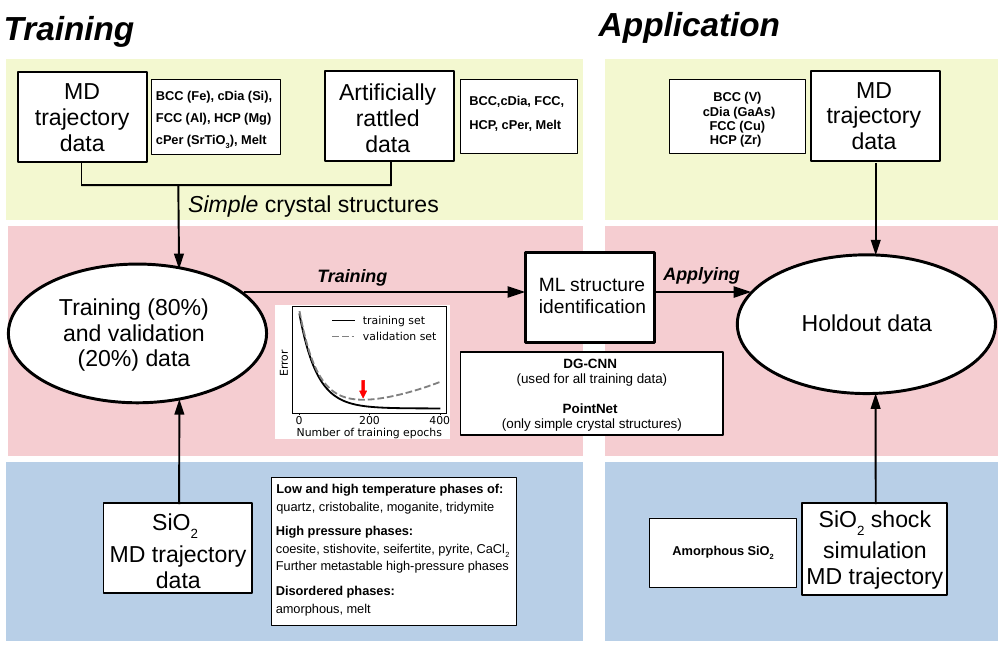}
    \caption{\textbf{Generation of training and holdout data.} Training data has been generated for \textit{simple} crystal structures using MD trajectories and artificially generated training data. While the MD trajectories have been generated with interatomic potentials corresponding to the elements shown in brackets, the artificially generated data has been created by rattling of structures. In case of \sio{} we generated training data for a large number of structures (Details see in the main text). In all cases we split the input data into training dataset and validation dataset. During the training the DG-CNN or PointNet are trained to the training data, however, the validation data is used as decision metric to pick a model. The fitted structure identification algorithms have been then applied to a holdout set. For the \textit{simple} crystal structures these have been MD trajectories of different elements, but with the same crystal structures. For silica we used a shock simulation to test our identification algorithm. The green background colors differentiate different parts of the workflow, specifically, data generation for \textit{simple} crystal structures (green), data generation for \sio{} (blue) and the ML workflow (red).}
    \label{fig:methods}
\end{figure}

Figure \ref{fig:methods} gives an overview of the methodological procedure in this work. Training and validation datasets are taken from from molecular dynamics simulations and artificially perturbed structures. Additionally, an independent holdout dataset is created using different chemical elements or molecular dynamics simulations based on alternative protocols, such as shock wave simulations. Further details are provided in the subsequent sections.

All MD simulations are carried out using \textsc{lammps} \cite{Thompson2022}. NNs are implemented in \textsc{pytorch} \cite{Paszke2019}. Data extraction and visualization is mostly implemented in \textsc{ovito} \cite{Stukowski2010} and aided by \textsc{numpy} \cite{VanDerWalt2011} and \textsc{scipy} \cite{Virtanen2020}. Processing is accelerated using \textsc{parallel} \cite{Tange2011a}.

\subsection{Training and validation data from molecular dynamics}

MD simulations were employed to generate training and validation data for various structure and interatomic potentials at temperatures. Each sample is made up of a different element and described by its own interatomic potential: BCC Fe \cite{Mendelev2003}, cDia Si \cite{Stillinger1985}, FCC Al \cite{Mendelev2008}, HCP Mg \cite{Sun2006}, cubic perovskite (cPer) {SrTiO\textsubscript{3}} \cite{Thomas2005}. Each sample was heated from \SI{1}{\kelvin} to \SI{20}{\kelvin} below the homogeneous bulk melting temperature in \SI{192}{\pico\second} (timestep length \SI{1}{\femto\second}). Afterwards each sample was heated to \SI{20}{\kelvin} above this melting temperature in \SI{20}{\pico\second} to ensure melting. This melt is subsequently heated to twice the melting temperature in \SI{192}{\pico\second}. During the two heating simulations \(2 \times 640\) snapshots with equidistant temporal spacing are stored. All simulations are run in the isobaric-isothermal (NPT) ensemble with fully periodic boundary conditions.

Training data for the \sio{} structures was generated using an ML potential, based on the atomic cluster expansion \cite{Erhard2023}. 
We used a time step of 1~fs, a thermostat damping coefficient of 100~fs and a barostat damping coefficient of 1000~fs. 
As input structures for the MD simulations we used supercells of 23 different crystalline silica phases.
These crystalline phase include stable silica polymorphs as well as the most common metastable ones. 
We added a comprehensive collection of metastable high-pressure silica phases, which were theoretically predicted or experimentally observed in compression or shock experiments. 
In detail, we have included the following phases $\alpha$-quartz, $\alpha$-cristobalite, $\alpha$-moganite, monoclinic tridymite, which all transform by heating to corresponding higher temperature polymorphs. 
Moreover, we added a list of stable and metastable high-pressure polymorphs: $P3_221$ \cite{badroTheoreticalStudyFivecoordinated1997}, $C2$ \cite{choudhuryInitioStudiesPhonon2006}, d-NiAs-type \cite{liuNewHighpressureModifications1978}, $I2/a$ \cite{tseHighpressureDensificationAmorphous1992}, NaTiF$_4$-type \cite{teterHighPressurePolymorphism1998}, $P2_1/c$ \cite{teterHighPressurePolymorphism1998}, $P3_2$ \cite{wentzcovitchNewPhasePressure1998}, $Pnc2$ \cite{svishchevOrthorhombicQuartzlikePolymorph1997}, SnO2-type \cite{teterHighPressurePolymorphism1998}, rosiaite-type \cite{otzenEvidenceRosiaitestructuredHighpressure2023}, coesite \cite{kirfelEndingP21Coesite1984a}, stishovite \cite{keskarStructuralPropertiesNine1992}, seifertite\cite{zhangInsituCrystalStructure2016} and pyrite-type\cite{kuwayamaPyriteTypeHighPressureForm2005}. 
At elevated pressures stishovite transforms dynamically into the CaCl$_2$ phase. 
Initially, we heated these structures from 10~K to 6000~K with a heating rate of 2~K/ps under NPT conditions allowing the change of all cell parameters including the angles. 
The applied pressures depend on the stability range of the phase. Phase transitions have been identified by analyzing the inner energy, volume and lattice parameters depending on temperature. 
Based on the identified phase transitions (see Supplementary Tab.~\ref{stab:structures}) we repeated the simulations with the same heating rate.
To receive structural data we dumped 640 snapshots for each temperature range in which one phase is stable. These dump files contain structures from 30~K above the lower transition temperature up to 30~K below the upper transition temperature. 

For tridymite, we could not differentiate between its various phases in detail. 
Therefore, we decided to differentiate only between low temperature (low symmetry) tridymite, where the angle of the cell is $\neq$ 90$^\circ$ degrees and high temperature tridymite, where all angles of the box are = 90$^\circ$. 
Melt structures are automatically generated once the original structures are heated to the point of melting. Additionally, some of the metastable high-pressure phases either transform into different crystalline structures or become amorphous before melting.
The data of these structures after transformation is not used for training. 
Amorphous snapshots are created by generating amorphous structural models at two different pressures (0~GPa and 200~GPa) with two different quench rates (10$^{12}$ and 10$^{13}$ K/s). 
These structural models are then compressed and decompressed with a (de)compression rate of 0.1 GPa/ps at constant temperature (T=500,1000,1500 and 2000~K). During this (de)compression we dumped again 640 snapshots. 

\subsection{Artificial training and validation data generation}

An alternative to MD simulations is artificial training data, where the atoms are displaced from the crystallographic positions. The advantage of this data generation scheme comes from the low cost since no MD simulations need to be performed. Due to this, there is also no requirement for an interatomic potential.

\begin{figure*}[tbp]
  \centering
  \includegraphics{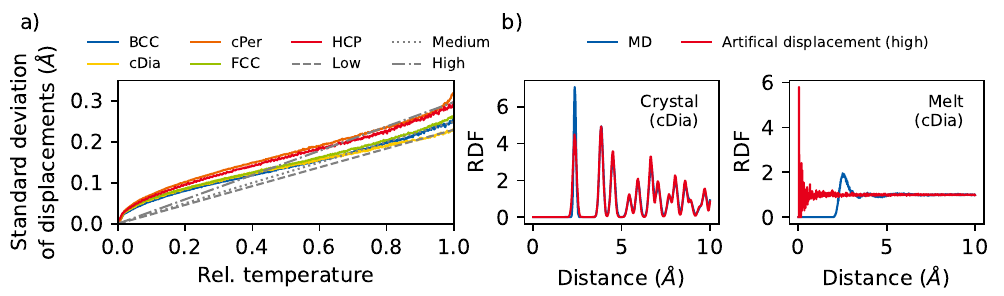}
  \caption{%
  \textbf{\textbf{Atomic displacements in MD and artificial data.} a)}~Standard deviation of the displacements from the ideal lattice sites in different crystal structures below (\textit{crystal}).
  \textbf{b)}~Values of the RDF for the cDia sample measured from MD simulations in the crystalline and liquid phase in comparison to the RDF obtained from the artificial displacement procedure outlined in the Methods section. Here, data for the greatest (\textit{high}) displacements are shown.
  \label{fig2}}
\end{figure*}

For generating artificial training data, we used Gaussian distributed displacements, since these agree well with the displacements in MD data (see \Cref{figS1}a). The amplitude of the displacements in dependence of the temperature in the MD simulations is depicted in \Cref{fig2}a. Since the MD training data contains 640 structure taken from different temperatures, we created 640 structures with different displacements corresponding to different 'artificial' temperatures. 
From the MD standard deviations of the displacements we derived four different relations between the these 'artificial' temperatures and the corresponding displacement amplitude. The first three relations are linear fits of the displacements observed in MD and are shown as \textit{low}, \textit{medium} and \textit{high} in the \Cref{fig2}a. In the last case the standard deviations of displacements are queried from the MD curves directly to get the most accurate displacement for each temperature. As shown in \Cref{fig2}b, the radial distribution function (RDF) of the crystalline structures agrees well between the MD and the artificially generated data.  
 To generate artificial data for the liquid phase, we introduced random displacements into the structures that are significantly larger than those for the crystalline phase, to facilitate the dissolution of crystalline order. The standard deviations of displacements we used are depicted in \Cref{figS1}b. The displacements are truncated at half the box size. To test the influence of the amplitude we also used different displacements modes for the liquid phase. In contrast to the crystalline case, there are significant deviations in the RDFs of the MD generated liquid phase and the artificially generated liquid phase (see \Cref{fig2}b). Due to the lack of repulsion in the artificially generated data, there are atoms with very small distances. However, this deviation from the MD data might be compensated later by distance normalization on the atomic environments prior to the processing using the NN.

\subsection{Holdout data generation}

The holdout dataset is a separate test dataset, which is generated from independent sources. While the original training and validation data stem from MD simulations using the same potential, now different elements/potentials are used in case of the simple systems. For \sio{}, we have made a simulation in which several phase transitions occur, including amorphous-crystalline and crystalline-crystalline phase transitions.

To generate a holdout set, the same approach outlined in the `training and validation data generation' section is taken. The interatomic potentials used are: BCC V \cite{Mendelev2007}, cDia GaAs \cite{Murdick2006}, FCC Cu \cite{Foiles1986}, HCP Zr \cite{Mendelev2007a}.

For \sio{} we additionally used another protocol for generating holdout data, namely shock wave simulations. Here, we used again the ML potential from Erhard \textit{et al.}~\cite{Erhard2023}. We used a time step of 1~fs and the \textsc{fix nphug} as it is implemented in \textsc{LAMMPS}. Within this algorithm time integration according to the Hugoniostat equations of motion is performed \cite{Ravelo2004}. As a target pressure we used a pressure of 50~GPa and allowed isostatic deformation of the cell. The temperature and pressure damping parameters were chosen to be 20~ps. The amorphous input structure for the shock simulations was created by melt-quench simulations with the quenching protocol from Ref.~\cite{Erhard2022}.

\subsection{Training and inference}

\begin{figure}[t]
  \centering
  \includegraphics{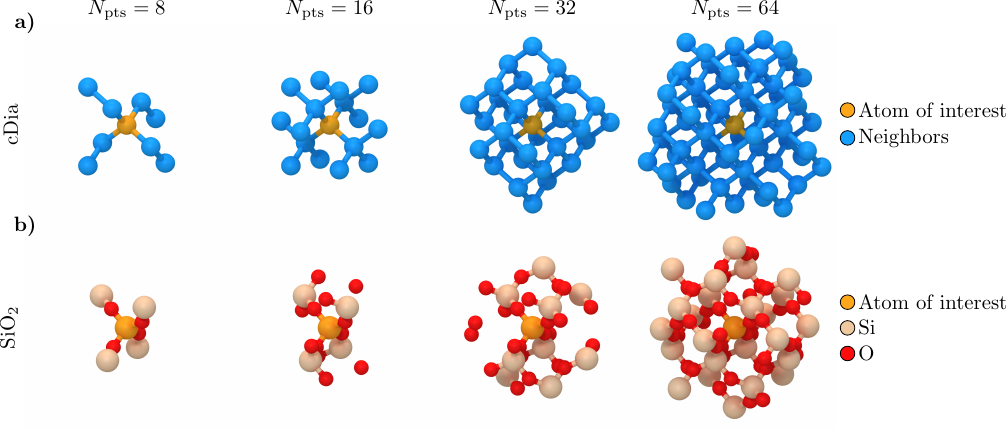}
  \caption{%
  \textbf{Input point clouds for structure identification.}
  Different environment sizes used as input into the NN structure classifier. Examples for both cubic diamond (cDia, \textbf{a}) and quartz (\sio{}, \textbf{b}) are given. The input environment can contain between \num{8} and \num{64} points (denoted as \(N_\text{pts}\)). The central atom, for which the structure type is determined based on this environment is always highlighted in yellow. The bonds shown in these snapshots are added for visual clarity, they are not used as input in the NNs. Only the atomic positions are used.
  \label{figS0}}
\end{figure}

From each sample for each phase considered, \(10^6\) environments were extracted. These contained, depending on the task between \(8+1\) and \(64+1\) atoms each. The \(+1\) refers to the central atom which we want to classify in a given step. \Cref{figS0} shows examples of these extracted environments for cDia and quartz at \SI{0}{\kelvin}. Each environment is shifted such that the target atom is placed at the origin. Afterwards, the whole group of atoms is scaled so that the nearest neighbor to the atom of interest has a distance of \SI{1}{\angstrom}. Lastly, the central atom is deleted as it is always in the same position and should not provide any structural information not implicitly contained in the positions of all neighboring atoms. Therefore, we end up with datasets of \(8 \times 3\) to \(64 \times 3\) values for each particle to be structurally identified. To accelerate training and inference all coordinates are stored and processed in single precision (32-bit) floating point representation.

For the solid phase, these \(10^6\)  environments were uniformly spread over the whole temperature range. For temperatures above the melting temperature, only structures where the cell volume was less than twice the volume at the melting temperature were considered. This filtering was necessary to exclude the gaseous phase from the training dataset.
Environments were extracted for random atoms throughout the sample. For samples containing more than one element, i.e., cPer, cDia, and \sio{}, each species was selected with the same probability. This was done to try to ensure the same classification accuracy for every species.

We use data augmentation during training to train rotation and point order invariant networks. Every time an environment is loaded into the network it is randomly rotated around the atom of interest and the order of points is shuffled.
During inference only the rescaling of the next-neighbour distance to \SI{1}{\angstrom} and deletion of the central atom are applied. Random rotation and point shuffling will not give any benefits here. 
The output of the ML algorithm is a classification score for each structure type, which is normailzed to a range of \num{0} to \num{1} using the softmax function. 
As discussed later on, this does not directly correspond to the corresponding accuracy. 
Nevertheless, we decided for the simple systems that scores below \num{0.5} will be labeled as \textit{Other} phase to show that no clear classification could be made. 
This notation is borrowed from the PTM method. 
However, since we defined a disordered amorphous class and a melt class for the \sio{} system, we did not define a \textit{Other} class phase there.

The \textsc{pytorch} implementation of PointNet \cite{Qi2016} is taken from user \textsc{fxia22} on \textsc{github} \footnote{\url{https://github.com/fxia22/pointnet.pytorch} (MIT license)}. The network was trained for \num{250} epochs on a total of \num{500000} environments evenly taken from the different crystalline and the liquid phase. This data is split into \SI{80}{\percent} training and \SI{20}{\percent} validation set. The batch size is set to \num{256}. For optimization, the Adams algorithm \cite{Kingma2015} with an initial learning rate set to \(10^{-3}\) is employed. The learning rate is reduced to a minimum of \(10^{-5}\) using a cosine annealing schedule \cite{Loshchilov2017}. Lastly, we activate the optional \textit{feature transform} setting in PointNet.

We use DG-CNN \cite{Wang2019} as implemented in the \textsc{pytorch} framework by \textsc{github} user \textsc{AnTao97} \footnote{\url{https://github.com/AnTao97/dgcnn.pytorch} (MIT license)}. Similar to PoinNet, training is run for \num{250} epochs on \num{500000} environments from each phase of interest. A \(80/20\) training/validation split is used. The learning rate of the Adam optimizer \cite{Kingma2015} is reduced based on cosine annealing \cite{Loshchilov2017} from \(10^{-3}\) in case of the \textit{simple} crystal structures and \(10^{-4}\) in case of \sio{} to \(10^{-5}\). Here, the batch size is reduced to \num{128}. DG-CNN comes with more parameters: we set the \textit{edge feature set size} to \num{8} and the \textit{max pooling output size} to \num{16}. Further, we deactivate \textit{dim9}. The dropout for the multilayer perceptron output layers is set to \SI{20}{\percent}. 

\section{Performance for simple crystal structures}

To test the applicability of NNs we perform preliminary studies on structures for which a well-established analytical classification algorithm exists: BCC, cDIA, cPer, FCC, and HCP. All of these, except for cPer, can currently be identified using CNA or PTM methods implemented in \textsc{ovito}, providing a solid benchmark for the NN classification. cPer was added as a crystal structure that cannot be identified automatically to show that NN methods can extend the scope of the existing algorithms.

\subsection{Benchmarking of PointNet and DG-CNN on simple crystal structures}

\begin{figure*}[tbp]
  \centering
  \includegraphics{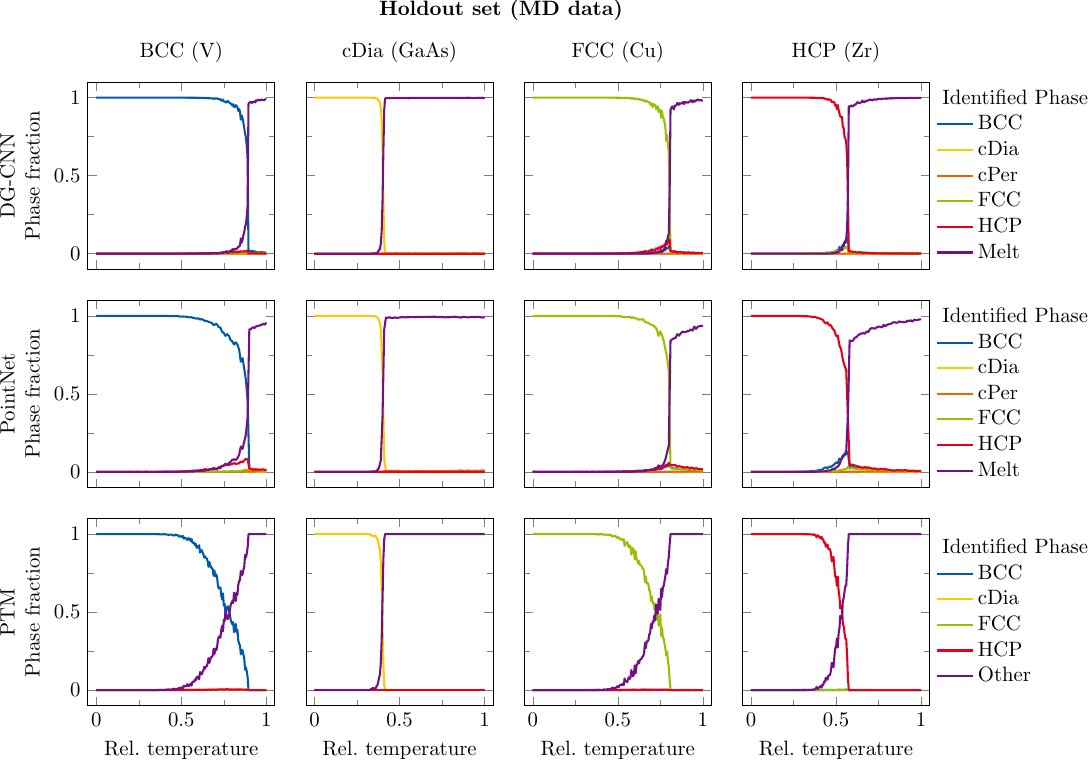}
  \caption{
    \textbf{Performance of PointNet and DG-CNN on the holdout set.}
    Comparison of the structure classification of the two different NN architectures (DG-CNN and PointNet) in comparison to the PTM method. Here classification of BCC, cDia, FCC, and HCP samples from the holdout set are compared. They are heated to an arbitrarily selected temperature above the homogeneous bulk melting temperature. Therefore, only a relative (rel.) temperature in reference to this maximum is given. Melting can be seen by the pronounced drop in crystalline phase. The NNs shown here are trained on MD trajectories, refer to the Methodology section for further details.
  \label{fig1}}
\end{figure*}

First, we will compare the PointNet and DG-CNN architectures using comparable hyperparameters and training data. Training is performed on \num{500000} input environments, containing 32 atoms each, obtained from MD trajectories. After training, PointNet and DG-CNN achieved average accuracies of \num{0.983} and \num{0.992} on the validation set, respectively. However, it is important to note that the validation data is obtained from the same MD simulation as the training data. Therefore, there may be some spurious correlations in this dataset which were learned by the NNs. To confirm that they indeed learned structural features, we now use the two trained networks to classify MD simulation data from the holdout set. This data was simulated using a different set of interatomic potentials, allowing us to test the transferability of the models.

\Cref{fig1} shows the fraction of atoms belonging to a given crystal structure for the two NNs in comparison to the PTM classification results. Comparison of the 4 different material systems reveals that the cDia structure can be identified most easily by all 3 methods. Here the transition from crystalline to molten (labeled \textit{Other} by PTM) phase can be seen most clearly and most sharply. For the other three systems, a more broad transition regime can be seen. This transition is most pronounced in the PTM results, while both PointNet and DG-CNN show a much sharper transition. This probably stems from the fact that they were both trained on low and high temperature MD trajectories. PTM on the other hand is based on the ground state crystal structure, which is then allowed to have some deviation from that ideal atomic arrangement. This degree of deviation is the key parameter in this algorithm, governing the degree of distortion up to which a crystal structure is recognized \cite{Larsen2016}. 
We observe that the two NN architectures show comparable classification results for the crystalline phases, even up to very high temperatures. There might be a slight edge in performance for the DG-CNN classifier most clearly visible in the BCC sample. However, the difference in both networks becomes most apparent in the classification accuracy of melt phase. Here, we can see that the DG-CNN usually identifies \SI{100}{\percent} of melt beyond the melting temperature. The PointNet, on the other hand, finds about \num{5} to \SI{10}{\percent} of non-melt structures, even beyond the melting temperature. From these results, we conclude that the usage in structure identification DG-CNN outperforms PointNet.

\subsection{DG-CNN: Comparsion of artifical and MD training data}

Based on the available literature on ML based structure identification methods two different approaches for training data generation can be taken. DeFever et al.\cite{DeFever2019} rely on MD trajectories for their training data. They performed their training and validation on MD simulations based on the same interatomic potential. Note that they performed training and validation on MD trajectories of the same interatomic potential without exploring extrapolation to other systems. 
Other studies \cite{Leitherer2021,Chung2022} use synthetic training data. This means that they apply random displacements to the ideal crystalline atomic positions to approximate the effects of temperature without costly simulations. 
Leitherer et al.\cite{Leitherer2021} apply uniform random displacements in five discrete displacement amplitudes between \SI{0.1}{\percent} and \SI{4}{\percent}. They also augment their dataset with missing atoms, something not considered here. Chung et al.\cite{Chung2022} also opt for random displacements uniformly distributed in a sphere with a radius of \SI{25}{\percent} of the nearest neighbor distance.
In this section, we compare the classification performance of DG-CNN trained on synthetic -- also called \textit{artificial} -- training data in the following, to NNs trained on MD trajectory data. Again, the performance is assessed based on a holdout set to exclude spurious correlations the network might have learned from the MD trajectories.

\begin{figure*}[tbp]
  \centering
  \includegraphics{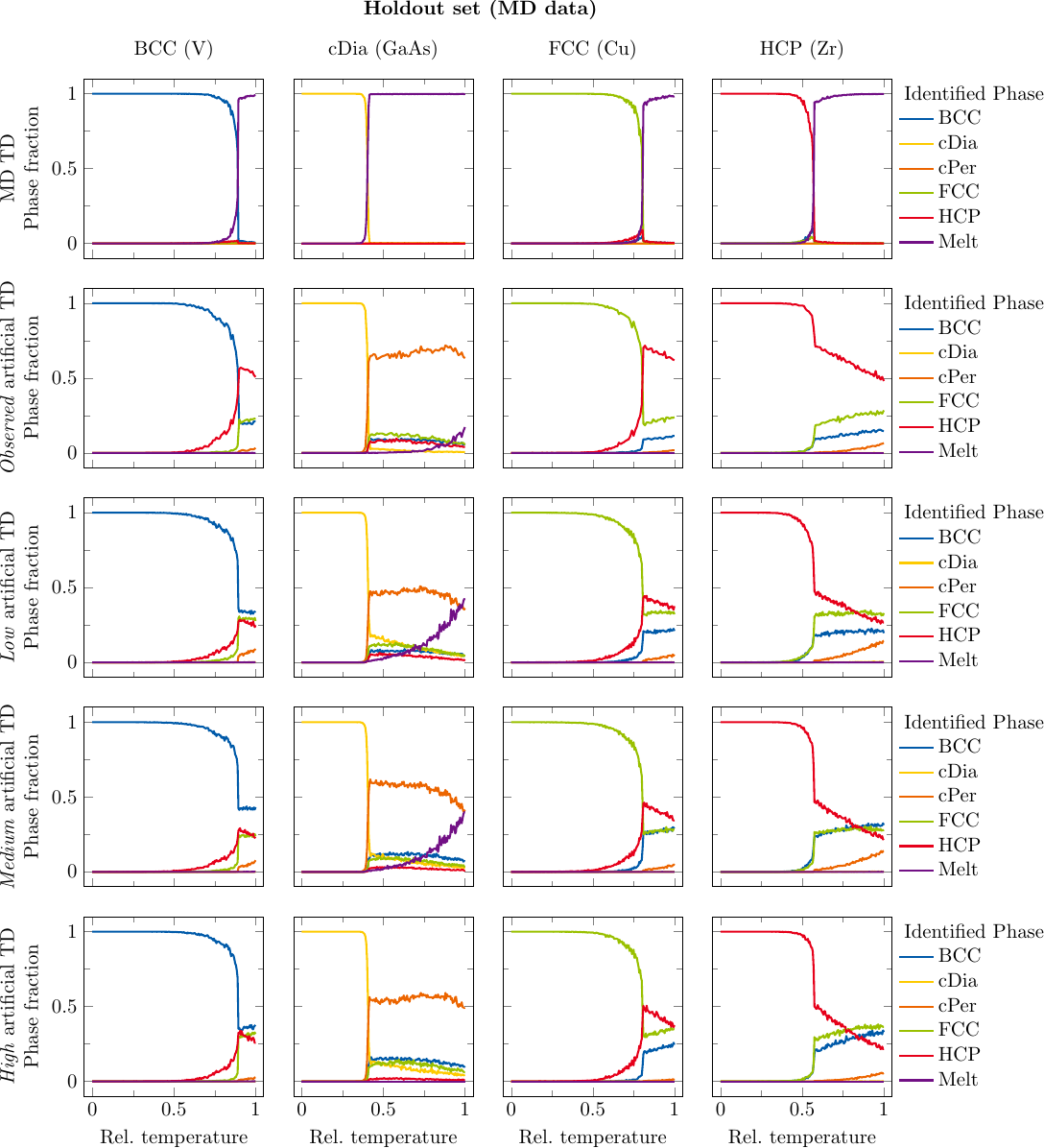}
  \caption{
  \textbf{Influence of MD and artificially generated training data} (TD)
  on the DG-CNN accuracy on the holdout dataset. In all cases, classification results for BCC, cDia, FCC, and HCP samples from the holdout set are shown. Each sample is heated to an arbitrarily selected temperature above the homogeneous bulk melting temperature. Therefore, only a relative (rel.) temperature in reference to this maximum is given. Melting can be seen by the pronounced drop in crystalline phase. See \Cref{fig2}a for a description of the different artificial training data generation parameters.
  \label{fig3}}
\end{figure*}

For the artificial training data we created four different datasets with different rattling amplitude called \textit{low}, \textit{medium}, \textit{high} and \textit{observed}, which base directly on the MD displacements (see Methods). We also created artificial melt structures with significantly higher displacements. Based on these artificial training datasets and the MD training data 5 different DG-CNNs are trained. The resulting structure classification are given in \Cref{fig3}. After training, the different networks depicted here achieve average classification accuracies of \num{0.992}, \num{0.988}, \num{0.997}, \num{0.996}, and \num{0.991} for the MD, \textit{observed}, \textit{low}, \textit{medium}, and \textit{high} training data generation methods, respectively. 
However, for applicability to real world classification data the performance on the holdout set is much more important. This is shown in \Cref{fig3} and support the interpretations from \Cref{fig2}b that the artificial training data cannot correctly capture the first nearest neighbor correlations in the melt and therefore is not able to identify the liquid phase correctly. Evidence for this can be seen for all DG-CNNs trained on the artificial training data where there is a pronounced drop in the main crystalline phase. However, instead of a strong increase in melt at the same temperature, these NNs predict the appearance of other crystalline phases. This indicates that the melt phase was not trained correctly leading to other crystalline phases appearing more correct to these NNs. Only the NN trained on MD trajectories gives the expected results. 
Interestingly the main crystalline phase is identified with approximately the same accuracy by all the five classifiers. 

Two main conclusions can be drawn from these results. First, for crystalline phases artificially generated training data appears to be reasonably suitable even up to the melting temperature, with only a slight edge in performance for the fully MD-trained NN. This supports previous works based on this approach \cite{Ziletti2018, Leitherer2021, Chung2022}. Moreover, the classification accuracy shows only a weak dependence on the displacement amplitude. Additionally, our approach extends these works by including the explicit identification of the liquid phase. The only other work also working with molten phase, trained on MD trajectories and obtained very high accuracies \cite{DeFever2019}. Again, showing results in-line with ours.
For now, if one wants to classify liquid phases, MD trajectories of melt are needed for training. However, a combination of artificial training data for crystalline phases and MD training data for the liquid phase might be a successful alternative to speed up the training process. Developing an algorithm for artificial training data generation for the melt probably involves some limitations on interatomic distances to avoid atoms being very close together as show in \Cref{fig2}. While this might already be sufficient, however, capturing the correlations leading to the first peak in the RDF might be necessary as well. 
Lastly, caution is needed regarding accuracy on the validation set. While all networks demonstrate excellent performance with accuracies above \num{0.98}, their application to the holdout set reveals that this data is not representative of the real-world scenarios, particularly for the liquid phase.

\subsection{DG-CNN: Impact of environment size for simple crystal structures}

\begin{figure*}[tbp]
  \centering
  \includegraphics{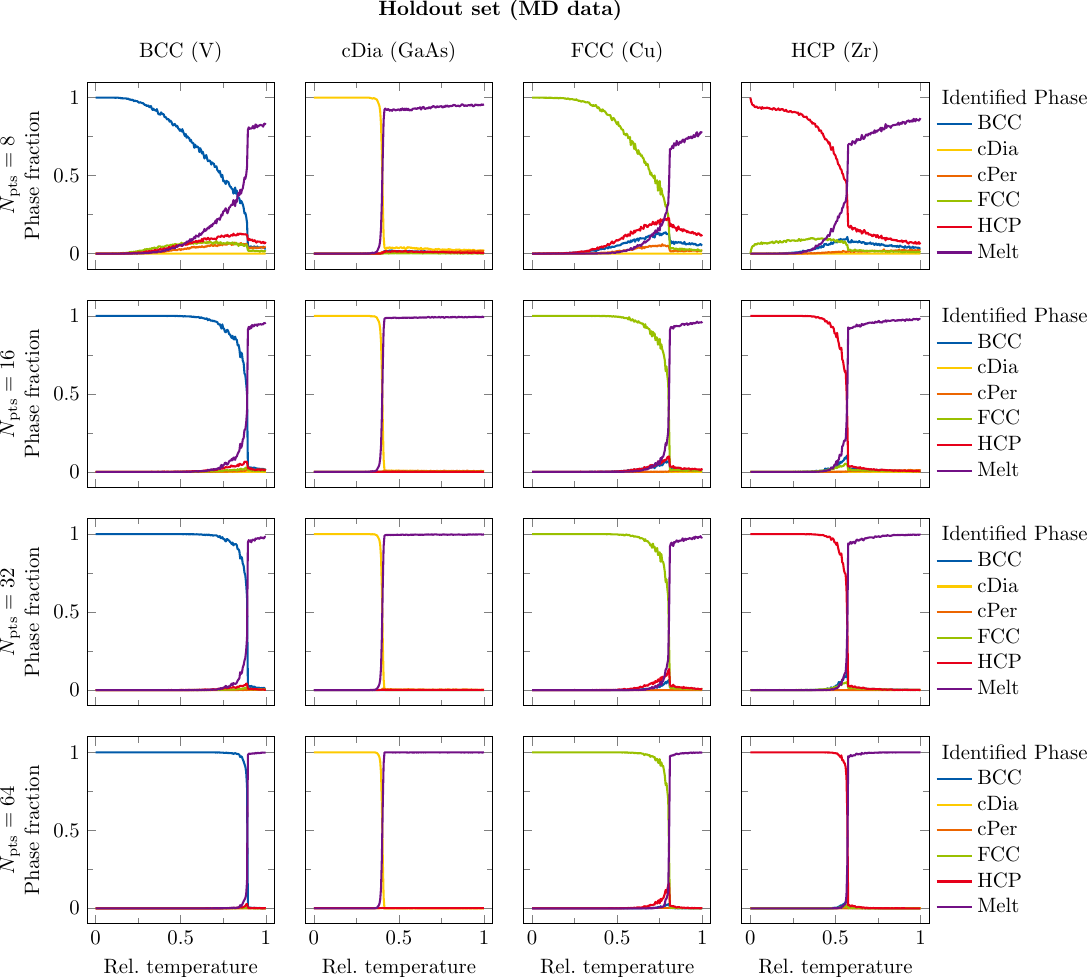}
  \caption{%
  \textbf{Impact of the enviroment size.}
  Crystal structures identified based on DG-CNNs trained using different input environment sizes \(N_\text{pts}\) from \num{8} to \num{64} atoms. In all cases, classification results for BCC, cDia, FCC, and HCP samples from a holdout dataset are shown. Each sample is heated to an arbitrarily selected temperature above the homogeneous bulk melting temperature. Therefore, only a relative (rel.) temperature in reference to this maximum temperature can be given. Melting can be seen by the pronounced drop in crystalline phase.
  \label{fig4}}
\end{figure*}

\begin{figure*}[tbp]
  \centering
  \includegraphics{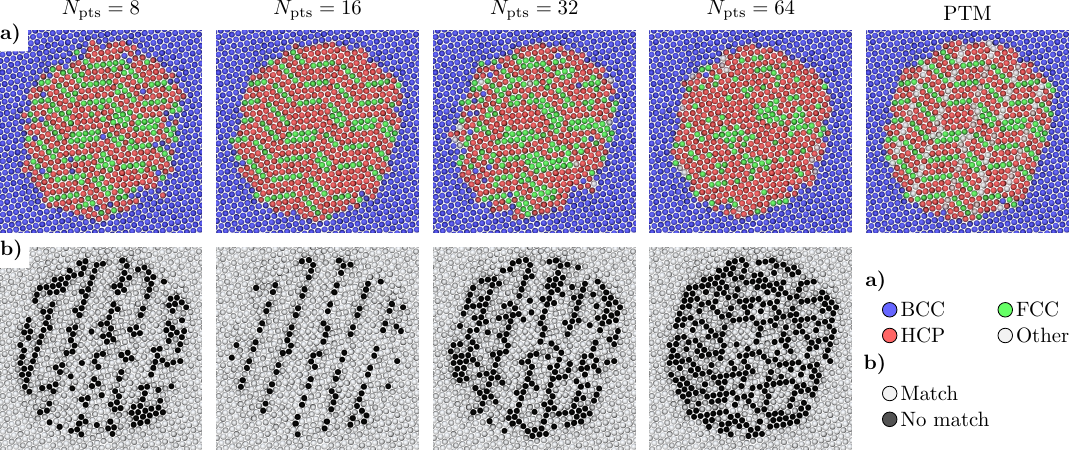}
  \caption{%
  \textbf{Spatial resolution dependence on environment size.}
  9R Cu precipitate in a BCC Fe matrix taken from \cite{Erhart2013}. The sample is used to assess the accuracy-resolution trade-off in the DG-CNN classifiers. Here larger chemical environments give higher accuracies on the single crystalline samples but have lower spatial resolution. For reference, the PTM classification result is given as well. PTM is a well established algorithm with good spatial resolution.
  \textbf{a)} Atoms color coded based on their respective structure types. 
  \text{b)} Differences between the DG-CNNs and the PTM classification are highlighted. Atoms with the same type in both methods are shown in white (\textit{match}) while atoms with differing classifications are shown in black (\textit{no match}).
  \label{fig5}}
\end{figure*}

So far, all structure identification results have been obtained on local environment sizes of \num{32} atoms. In this last benchmark we will systematically vary the local environment size \(N_\text{pts}\) from \num{8} to \num{64} atoms and see the effect on the classification accuracy. However, when investigating the environment size the corresponding accuracy of the NN is not the only relevant parameter. A larger environment means that the structural information is extracted from a larger volume and therefore gives a lower spatial resolution. On the other hand small environments not only give greater spatial resolution at the cost of more noise, they also increase the classification speed, as processing fewer points is faster.

In the following, we will compare DG-CNN networks trained on \num{500000} environments of varying sizes. Here the MD trajectories are used as training data. After training the networks reach average validation accuracies of: \num{0.852} (\(N_\text{pts}\) = 8), \num{0.983} (\(N_\text{pts}\) = 16), \num{0.992} (\(N_\text{pts}\) = 32), and \num{0.998} (\(N_\text{pts}\) = 64). Similar to the previous characterization, these new NNs are now benchmarked on a holdout dataset prepared using different interatomic potentials. The resulting structure identification results are given in \Cref{fig4}. 
Here we can see the trends from the validation accuracy replicated on the holdout set. As the environment size is increased, the phase transition gets sharper and the separation between solid and liquid phase is more defined. Based on these classification results, it can be concluded that an environment size of \num{8} atoms is too small for accurate classification once there is any thermal noise on the atomic position data. This could be caused by the fact that all structures considered here have more than \num{8} atoms in their nearest neighbor shell. 
Once larger environments of \num{16} and \num{32} atoms are considered, classification accuracy becomes saturated. Lastly, the largest environments, \num{64} atoms, lead to a more defined phase transition.

From the previously shown data one might presume that larger environments are the way to go in structure identification, as they lead to a higher accuracy and more defined phase transitions. 
However, there is another trade-off that needs to be considered. Increasing the radius around the central atom that is considered for classification automatically leads to a reduction in spatial resolution.
To assess this effect, we use the different trained classifiers to characterize a 9R Cu precipitate in a BCC Fe matrix. The sample is taken from Erhart et al.\cite{Erhart2013} and has been used for benchmarking of structure identification algorithms before \cite{Stukowski2012a}.
The resulting classification results compared to those obtained from PTM (RMSD of \num{0.1}) are shown in \Cref{fig5}a. For clarification \Cref{fig5}b highlights all atoms in the slice where DG-CNN and PTM do not agree with each other. Here we can see the aforementioned effects. The NN trained on \num{8} atoms shows the most noisy classification result where some atoms inside the precipitate are identified as BCC phase, even though it should be FCC and HCP. Additionally, the bottom right and top left sections of the particle seems to be classified as BCC matrix. 
This changes for the \num{16} atom environment where we get an almost perfect match of PTM and NN classification result. The main difference is that the PTM algorithm labels the planes where differently oriented twins meet as \textit{Other} while the NN identifies these mostly as HCP phase.
Once we get into the larger input sizes of \num{32} and \num{64} atoms it can be seen that the NNs cannot separate the narrow bands of HCP and FCC phases in the structurally complex particle. Here the larger networks mostly identify larger clusters of either FCC or HCP instead of the discrete layers.

The very good performance of the \(N_\text{pts} = 16\) network in this test case comes as a bit of a surprise. The 9R particle shows many interfaces between phases, however, the NN has never been trained on training data containing phase boundaries. Therefore, this suggests that the network is able to extrapolate from the pure phase information to more complicated multiphase samples. This we tested additionally in Fig. \ref{fig:hea_polycrystal}, where  we present  the structure identification result of an high-entropy polycrystal in both its undeformed and deformed states, which is taken from Ref. \citenum{nag2024}. The structure, comprising various atom types, grain boundaries, and stacking faults, poses a significant challenge for structure identification algorithms. Both PTM and DG-CNN \(N_\text{pts} = 16\) successfully identify the grains as FCC, with grain boundaries clearly distinguishable from the grain interior. Similar to the 9R Cu precipitate scenario, atoms at the grain boundary are classified as \textit{Other} by PTM, but are often identified as the HCP phase by the DG-CNN. When observing the polycrystal under tension, we note that both methods effectively reveal the stacking faults, showing good agreement.

\section{Case study of a structurally complex material}

In this chapter, we apply the DG-CNN to the structurally complex \sio{} system. The polymorphism of \sio{} makes it a challenging system for structure identification \cite{heaneySilicaPhysicalBehavior1994a}. Additionally, some phases experience solid-solid phase transitions when subjected to temperature changes even in the time scale of MD simulations \cite{prydeSequencePhaseTransitions1998,ramanAvTransformationQuartz1940,leadbetterTransitionCristobalitePhases1976}. Moreover, the amorphous silica phase is essential for many applications and should be therefore also included. Under high pressure silica has been observed in a wide variety of structures \cite{badroTheoreticalStudyFivecoordinated1997,choudhuryInitioStudiesPhonon2006,liuNewHighpressureModifications1978,tseHighpressureDensificationAmorphous1992,teterHighPressurePolymorphism1998,wentzcovitchNewPhasePressure1998,svishchevOrthorhombicQuartzlikePolymorph1997,otzenEvidenceRosiaitestructuredHighpressure2023}. For understanding the outcome of shock and compression simulations it would therefore highly advantageous if these structures could be easily identified. 
For training the DG-CNN we generated a database containing these types of structures using a ML potential \cite{Erhard2023} in MD simulations. Details about the choosen structures and MD protocols used can be found in the Methods section and in Tab. \ref{stab:structures}.

\subsection{Crystalline polytypes and structure recognition}

\begin{figure*}[tbp]
  \centering
  \includegraphics[width=17cm]{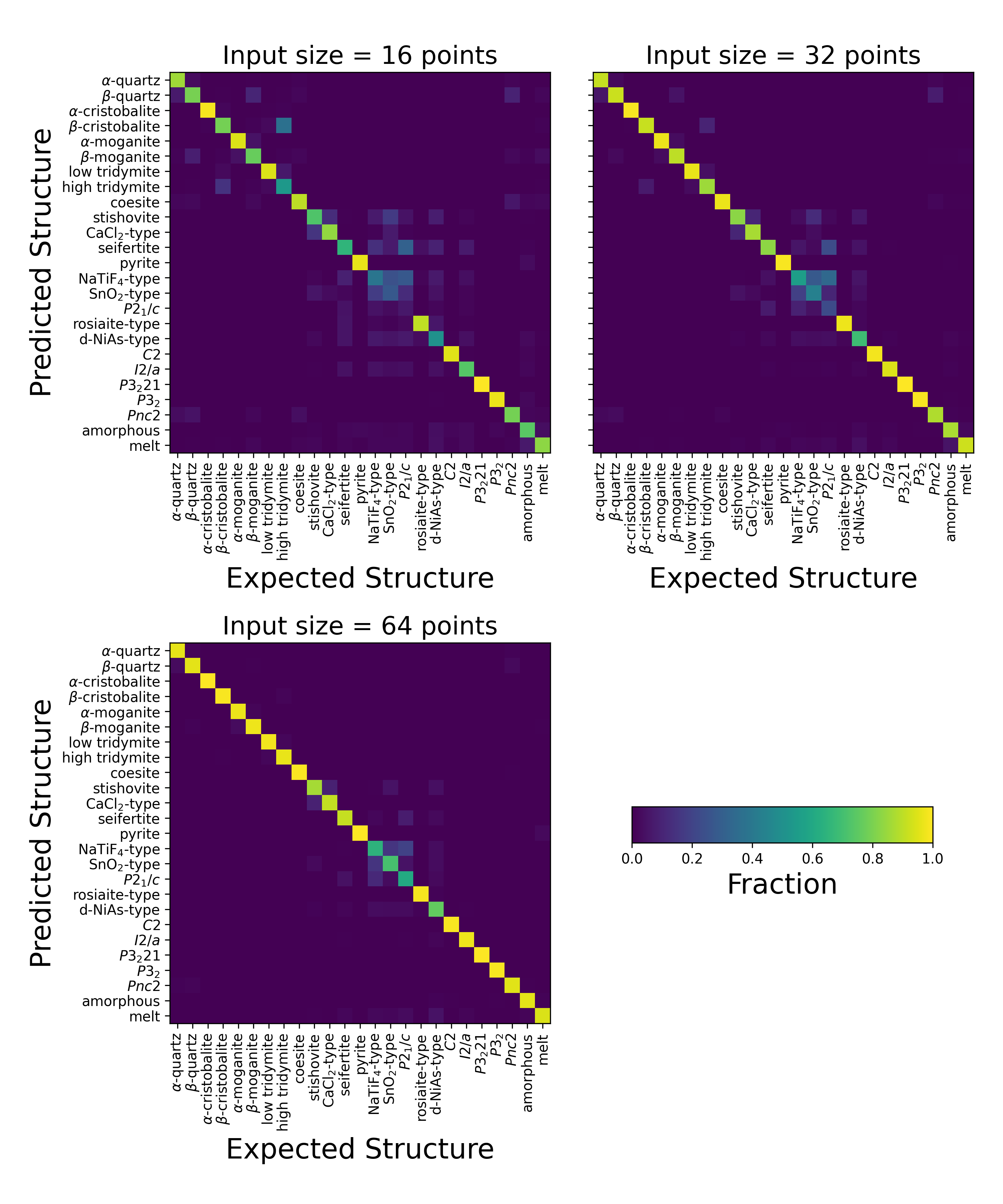}
  \caption{%
  \textbf{DG-CNN performance for silica.}
  Confusion matrix for the \sio{} system showing the expected and predicted crystal structures for classifier DG-CNNs trained on MD trajectories. The confusion matrix show the evaluation behavior on the full MD trajectories, from which also the training and validation set have been extracted. Each network was trained on a different number of input points \(N_\text{pts}\). A greater number of points corresponds to a larger environment being taken into account to classify the central atom. 
  \label{fig6}}
\end{figure*}

While we determined \num{16} to be the optimal value of environment size for the \textit{simpler} crystal structures (cf.\ \Cref{fig4,fig5}) the much larger structural groups in the \sio{} phases might require a different number of atoms as input to the NN. Therefore, we benchmark environment sizes \(N_\text{pts}\) of \num{16}, \num{32}, and \num{64} particles. These reach average accuracies of \num{0.75}, \num{0.85}, and \num{0.92} on the validation data, respectively.
As already discussed previously the environment size comes with different trade-offs. Larger environment allow for a more accurate structure identification at higher computational cost and lower spatial resolution.

\Cref{fig6} shows the confusion matrix of the different \sio{} phases considered in this work. Compared to the results for the \textit{simpler} crystal structures, these have not been strictly calculated from a holdout dataset as there is no other material system showing the same phase diagram. However, the confusion matrix is calculated on the full MD trajectories from which the training and validation data has been extracted. We note that training and validation data account for less than 3\% of these MD trajectories, therefore, there are many previously unseen environments. 
The confusion matrices show the expected crystal structure against the one predicted by DG-CNN classification. Here we can see that for most structures the DG-CNN correctly identifies the crystal structure. These correct predictions can be found on the main diagonal of the plot. 
However, if we have a closer look, at the off-diagonal elements, we can see that there also several problematic structures where miss-classification occur.

First, we have a look at the difficulties differentiating between $\beta$-cristobalite and high temperature tridymite. Both structures are similar. 
In fact, the silicon sublattices of $\beta$-cristobalite is a cubic diamond lattice, while the sublattices of tridymite is a hexagonal diamond lattice. 
This makes it difficult for the DG-CNN with 16 input points to differentiate between both structures, however, with 32 and 64 input points the performance is significantly improved. 
Similar behavior is observed for low temperature and high temperature tridymite. Both structures convert dynamically into each other in the MD simulaton. While high temperature tridymite has a higher symmetry than the low temperature tridymite structure, there are no bonds broken during the phase transition. 

Unlike the previous cases, the DG-CNN finds it challenging to differentiate between stishovite and the CaCl$_2$-type, regardless of the number of input atoms.
The reason for this is that both structures are extremely similar and differ only by small distortion of the lattice parameters. 
While stishovite is tetragonal with a=b, CaCl$_2$-type silica is orthorhombic with a$\neq$b. 
Since in MD simulations both phases transform dynamically into each other, it might be that the training data is partially overlapping. 
Due to thermal fluctuations it might be that atomic environments are in some case shortly identical for both structures. 
Consequently, for a perfect differentiation clean input data would be needed.
However, it is not clear whether it is possible to get this data.
Only after this, it would be clear whether more input points could improve the classification process. 

The most challenging group of structures are NaTiF$_4$-, SnO$_2$- and $P2_1/c$-type silica as it can be seen in the confusion matrix. All of these structures have a HCP oxygen sublattice. This is similar to stishovite (distorted HCP), CaCl$_2$-type, seifertite, rosiaite-type and d-NiAs-type silica. However, additionally, the ordering of the silicon atoms in the octahedral interstices is very similar in all cases. 
In the SnO$_2$-type silica, the silicon atoms are arranged in 4$\times$4 zigzag chains. For the NaTiF$_4$-type silica, silicon is arranged along of 3$\times$3 zigzag chains and for the $P2_1/c$-silica, it is arranged along 3$\times$2 zigzag chains. 
A more in depth structural analysis of the structures can be found in Ref.~\citenum{teterHighPressurePolymorphism1998}. 
This similar zigzag ordering between all structure requires a large environment around the atoms to identify the structure uniquely.
That is also the reason for the poor performance of the DG-CNNs with input points of 16 and 32 and the much better performance of the 64 input point DG-CNN.

\subsection{Shock loading simulations of amorphous structures and structure recognition}

\begin{figure*}[tbp]
  \centering
  \includegraphics[width=17cm]{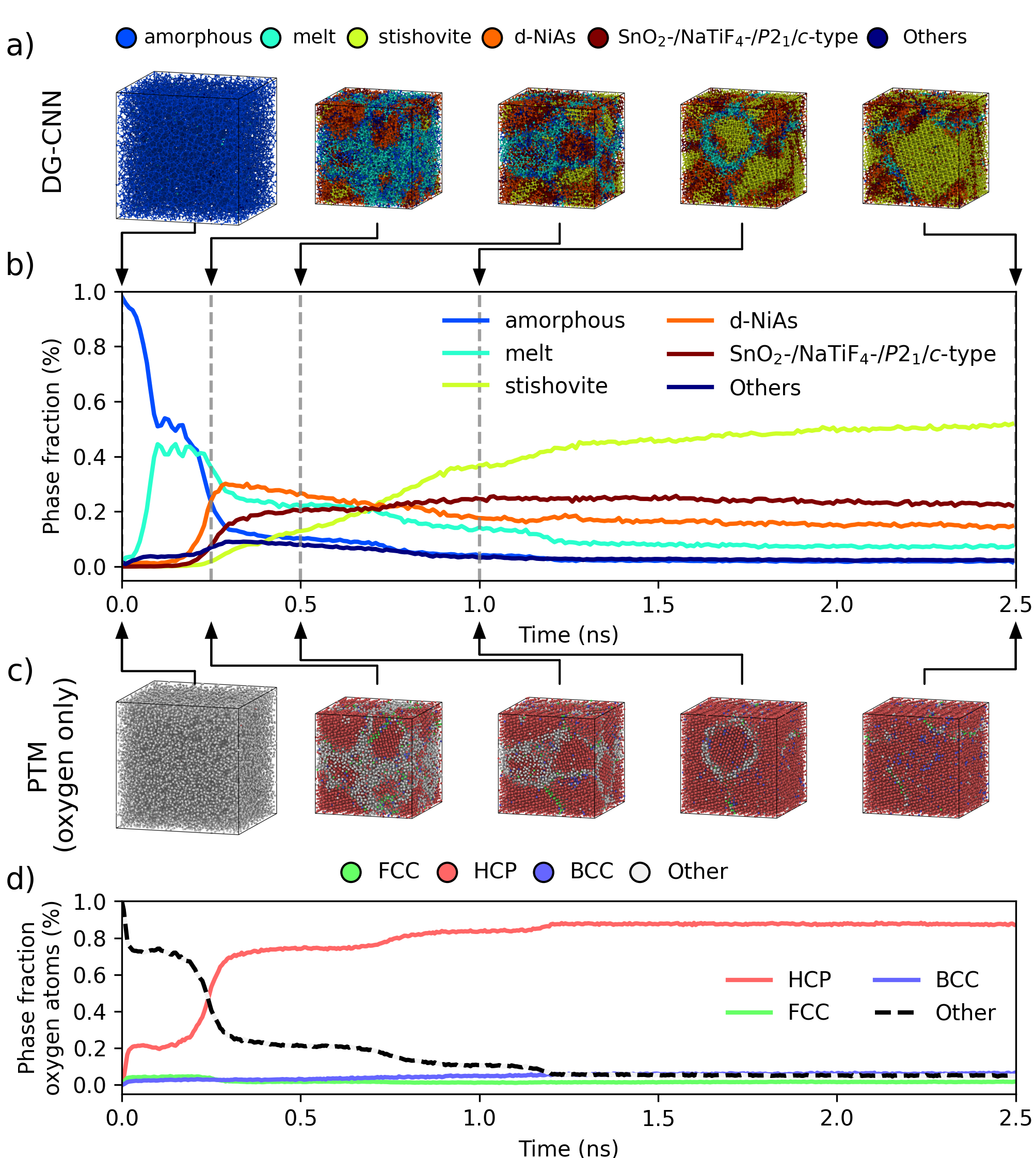}
  \caption{%
  \textbf{Shock compression of amorphous silica.}
  a) Structural evolution in an initially amorphous \sio{} sample which is subjected to shock loading. The DG-CNN recognised structure of each atom is marked by color coding. We used 64 input points. The structure class of SnO$_2$-, NaTiF$_4$- and $P2_1/c$-type silica is summarized as one. All other structure types are summarized as \textit{Others}. (b) Shows the phase fractions during this transformation.  c) Shows for some snapshots the classification of the oxygen sublattice of these structures by PTM (RMSD = 0.2), while d) shows the detailed analysis of the PTM phases over time. For comparison, classification results for NNs trained on environment sizes of \num{16} and \num{32} atoms can be found in \Cref{figS3,figS4}, respectively.
  \label{fig8}}
\end{figure*}

As a case study, we start with an amorphous \sio{} sample and subject it to a shock. Experimentally, it was observed that amorphous silica is crystallizing to stishovite under shocks equivalent to pressure above 36~GPa \cite{Tracy2018}. Also, earlier molecular dynamics simulations with simple classical interatomic potentials have shown similar behavior \cite{Shen2016}. Therefore, one would expect that the local environment of silicon changes from a fourfold to sixfold coordinated environment. At the same time, crystallization and grain growth can be expected. 

\Cref{fig8} shows snapshots of this shock simulations as well as the phase fractions determined by the DG-CNN taking \num{64} atoms as input environment size. The phase of each atom is identified according to the highest score from the DG-CNN.  

We can see that the initial amorphous sample is mainly classified as amorphous and a small amount as melt. 
Since both structure types are related, this is not surprising. 
When the amorphous sample is shock loaded, the DG-CNN predicted amount of melt is initially strongly increasing. 
This goes along with a strong increase in temperature, which in fact melt parts of the structure. 
After the initial partial melting of the structure, there is a strong increase in the amount of d-NiAs phase. Similar to stishovite and other phases, this phase also features an HCP oxygen sublattice.
However, in contrast to stishovite, SnO$2$-type, NaTiF$_4$-type and $P2_1/c$-type silica the silicon atoms are not ordered, but randomly distributed across the octahedral intersitial sites. 
This result agrees also well with results from a PTM analysis used only on the oxygen sublattice, which is shown in \Cref{fig8}c. 
These result show that at this point already large parts of the oxygen lattice are arranged in the HCP structure. 

After reaching a peak, the proportion of d-NiAs-type silica is decreasing again. This is caused by an ordering of the silicon atoms over the time changing the structure to more ordered structures like SnO$_2$/NaTiF$_4$/$P2_1/c$-type silica and especially stishovite. After 2.5~ns simulation time stishovite is the dominating polymorph with a phase fraction of roughly 50\%. This agrees well with the final state after the shock observed in experiment \cite{Tracy2018,gleasonUltrafastVisualizationCrystallization2015}.

In summary, our results suggest, that stishovite is not directly crystallizing out of the amorphous phase under shock. Instead, first a HCP sublattice is formed with randomly distributed silicon atoms (d-NiAs phase). Only later, these silicon atoms arrange in an ordered way, leading to the occurrence of stishovite.

\section{Discussion}

\subsection{Descriptor selection}

Inspired by the previous work of DeFever et al.\cite{DeFever2019} we decided to not use any derived descriptors but instead work with the atomic positions directly. This has the benefit that no further, time-consuming data transformation are needed prior to the actual classification work.

The main downside of this method is that SOAP and the like are inherently rotation and order invariant. Meaning that they can be used immediately as input for classification algorithms. The NN, on the other hand, has some regularization built in which should compensate for the ordering of input points and their rotation. However, in practice we found that it helps the NN learn and extrapolate to unseen data if the points in the training data are shuffled and rotated randomly each time they are shown to the NN.

DeFever et al.\cite{DeFever2019} train their PointNet on atomic environments taken from MD trajectory data. They use a fixed cutoff radius around the central atom they want to classify. This group of atoms is subsequently shifted to the coordinate origin and the nearest neighbor distances are scaled to some normalized distance.

However, since the NN expects a constant number of input points they have to either add dummy atoms at the origin or remove the furthest atoms with in their subvolume until the correct number of atoms is reached. 

This approach seems arbitrary and the effects of padding / deleting atoms on the classification accuracy is not explored further in their work. Therefore, we decide to always select a constant number of nearest neighbors of the atom that will be classified. This approach will fail for very small non-periodic cells, however, this should rarely be limiting in real world use. Moreover, we exclude the central atom (i.e. the one which has its environment classified). It will always be shifted to the coordinate origin and therefore does not contribute meaningful information to the environment. This gives us one additional atom to work with during classification.

There has also been a debate in literature over the right training data. While DeFever et al.\cite{DeFever2019} use MD training data to great effect, others \cite{Ziletti2018,Leitherer2021,Chung2022} rely on artificially generated training structures obtained from rattling of ideal crystal structures. While the latter are much easier to generate the former capture much for of the physics and correlated motion present in a real material. 
Our findings on \textit{simpler} crystal structures (\Cref{fig3}) show that this artificial generation of training data can work really well. However, we were not able to replicate this success for liquid phases. For now, we recommend validating networks trained on artificial training data against real MD trajectories to ensure that the synthetic data is representative of the real world. Moreover, new methods might be needed to generate good artificial training data for complicated systems like \sio{} or melt.

\subsection{Elemental information}

DG-CNN can include additional information about each point during training and inference. In computer vision task this might be the color of each point. Here, one could include the chemical information of the atoms to (presumably) increase classification accuracies. Another benefit of this approach is that the elemental information could be used to classify not only the crystal structure but also ordering in an alloy. Something that is already possible using PTM.
We decided against this approach because it also brings a set of problems. There would need to be some reduced representation of the chemical elements so that for example FCC Ni and FCC Cu can be identified as FCC structure by the same NN without explicitly training on both. Moreover, while one would like to separate for example an \(\text{L}1_0\) ordered FCC sample from an elemental one, a random solid solution of many elements on the FCC lattice should probably just be labeled FCC. Therefore, the input / output representation of the element types is quite challenging.

\begin{figure*}[tbp]
  \centering
  \includegraphics{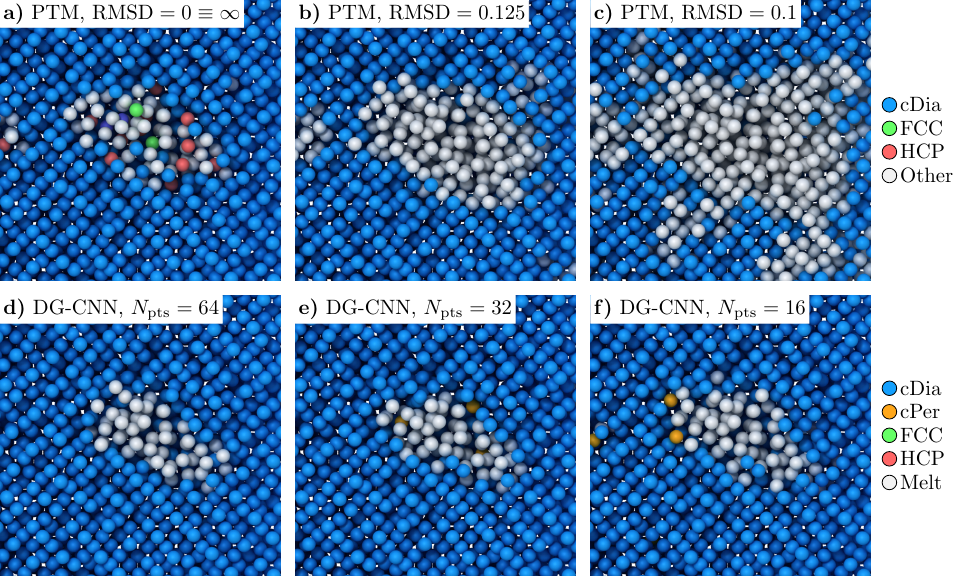}
  \caption{%
  \textbf{Ground truth problem.}
  Snapshots extracted from a cDia sample at temperatures very close to the bulk homogeneous melting temperature. A group of locally molten atoms in the still solid matrix can be seen in the center of each snapshot.
  \textbf{a-c} Atoms color coded based on the PTM structure identification result for different values of RMSD.
  \textbf{d-f} Structure classification from the DG-CNN classifier trained on MD data with different numbers of atoms \(N_\text{pts}\) included in the local environment used as NN input.
  \label{fig10}}
\end{figure*}

\subsection{Ground truth problem}

There is one last problem that needs to be addressed here. Namely, what is the ground truth for the crystal structure in simulated systems. \Cref{fig10} shows the same region in a cDia sample (from the holdout set) very close to the melting temperature. Note, that cDia was selected as sample because it shows the cleanest separation of crystalline and liquid phase for all identification methods. See \Cref{fig1,fig3} for details. 
The snapshot features a still crystalline matrix with a central region of homogeneous melt nucleation. The atoms in this figure are color coded based on PTM structure identification with different RMSD values in comparison to structure identification results from the MD trained DG-CNN with different input sizes (cf.\ \Cref{fig4}). The RMSD setting determines the allowed error between the ideal templates and the observed atomic environment in the PTM analysis \cite{Larsen2016} with greater RMSD values biasing the structure identification towards crystalline phase. As discussed before, a similar effect can be observed for the NN classifier where larger input sizes tend towards the identification of crystalline phase.

Without discussing the actual structure identification results we can see that the same region of crystal can lead to vastly different results. In \Cref{fig10}c many regions that look crystalline by eye appear as \textit{Other}, i.e.\ unidentifiable, phase. An RMSD of \num{0.125} seems to be a good compromise, however, it is unsatisfying that there is some arbitrary parameter which can so drastically change the structure identification result without a clear way of determining a \textit{good} value.

In contrast, the NN does not have any user tunable parameter. Here, the classification results and resolution are imposed during training of the model. The choices are, however, similarly arbitrary. Interestingly, the NN is able to characterize this two phase region in the sample, even though, the training data contains only single phase materials.

Similar decisions can also be found in other structure identification methods in literature. The \textsc{arise} framework identifies crystal structures where the atoms are displaced by up to \SI{10}{\percent} of the interatomic distance or up to \SI{20}{\percent} of atoms have been removed with almost perfect accuracy as the corresponding crystalline phase \cite{Leitherer2021}. Similarly, the \textsc{dc3} method is trained on structures where the atoms are displaced up to \SI{25}{\percent} of the interatomic distance \cite{Chung2022}. 

For reference, a material with \SI{7.7}{\percent} bulk vacancies will spontaneously collapse into liquid phase \cite{Fecht1992} and from \Cref{figS1} it can be seen that the assumption of random uniform atomic displacements is also far from the real material behavior.

This also poses the philosophical question of when a crystal is still classified as such, whether an atom deviating by twenty percent from its equilibrium position is still considered crystalline, or if a crystal with one-fifth of its atoms missing is still considered a crystal.
As long as these questions remain unanswered, crystal structure identification will continue to rely on decisions made by the developers and users of the code, emphasizing the necessity for justifying these decisions in each specific case.

\subsection{General remarks about NNs for structure classification}

The hyperparameters used in this work are basically hand tuned by the authors and seem to perform quite well. To validate that these parameters are reasonable we spent time optimizing them using the \textsc{optuna} library \cite{Akiba2019}. However, this process came at much higher computational cost for comparatively small accuracy improvements. Therefore, we did not continue this process. Especially, since we were able to show in this work that NN performance is much more limited by training data and descriptor choices. Two factors that cannot be addressed by any NN hyperparameter optimization library. Once clearly labeled training data, possibly including interfaces and defects, has become available, extensive hyperparameter tuning will make a lot more sense.

Lastly, the implementation of a \textit{nothing} class should be considered. The current network has no way of labeling fully unknown data. Currently, it always has to chose from the trained collection of classes. Chung et al.\cite{Chung2022} resolve this issue by implementing pre and post filtering of the classification data to recognize both amorphous and crystalline atoms of unknown crystal structure. An approach that might also proof fruitful for DG-CNN based classifiers as well. This could also aid in the classification of GBs by the network without explicit training on GBs as the GB structures might be closer to a \textit{nothing} class than their parent crystal structure or a melt or amorphous phase. 

\section{Summary}

DG-CNNs trained on MD trajectories and synthetic training data can be used to identify crystal structures. Here we found that while synthetic data is well suited for crystalline phases in \textit{simple} systems, MD training data gives much better classification results for melt. Moreover, for complicated systems like \sio{} using MD training data is unavoidable.

The environment size used for structure identification is of vital importance, as it governs the accuracy against resolution trade-off. For the \textit{simpler} systems much smaller environments of 16 points may be used compared to the complex \sio{} system, which needs even 64 points for an accurate description.

Here we show the first structure classifier for \num{25} different phases of \sio{}. This DG-CNN trained to identify the different \sio{} phases was used to classify the solid-solid phase transitions during shock loading of an initially amorphous sample. By this we had been able to achieve a deeper understanding of the crystallisation process of stishovite over an intermediated d-NiAs structure. 

In future, this approach can be easily extended to other systems with complex crystal structures. Furthermore, selecting appropriate training data makes it straightforward to expand the algorithm's capabilities to identify and classify defects in simulations. 

\section{Code and data availability}

We have made our DG-CNN implementation and surrounding code available as an \textsc{ovito} \textit{python script modifier}. Both this modifier and the neural networks trained in this work are available under the name \textsc{mlsi} (Machine Learning Structure Identifier) at \url{https://github.com/nnn911/MLSI}. This codebase may also be taken as a reference implementation of other \textsc{pytorch} or \textsc{tensorflow} NNs into \textsc{ovito} to process atomistic data and visualize the results.

Training data for both the \textit{simple} and \sio{} structures are available from \texttt{TUdatalib} at \url{https://doi.org/10.48328/tudatalib-1394}. 
Additionally, data from the MD holdout set and the \sio{} shockwave simulation will be made available at the same repository. The data loader and model implementation included in the \textsc{mlsi} repository \url{https://github.com/nnn911/MLSI} can be used as a basis to set up your own \textsc{pytorch} training environment. The 9R precipitate sample shown in \Cref{fig5} is not included as the rights lie with the original authors \cite{Erhart2013}.

\section{Acknowledgment}

The authors thank Shankha Nag for providing the high entropy alloy polycrystalline structures.
L.C.E. thanks Niklas Leimeroth and Christoph Otzen for helpful discussion. 
D.U. gratefully acknowledges funding by the NHR4CES research project as part of SDL `Materials'. 
The research was supported by the Bundesministerium für Bildung und Forschung
(BMBF) within the project FESTBATT under Grant No. 03XP0174A. 
The authors gratefully acknowledge the computing time provided to them on the high-performance computer Lichtenberg at the NHR Centers NHR4CES at TU Darmstadt. This is funded by the Federal Ministry of Education and Research, and the state governments participating on the basis of the resolutions of the GWK for national high performance computing at universities. This work was performed on the HoreKa supercomputer funded by the Ministry of Science, Research and the Arts Baden-Württemberg and by the Federal Ministry of Education and Research.

\section{Conflict of interest}
D.U. works at the time of submission for the OVITO GmbH. The research project was conceived and completed during their employment at TU Darmstadt.

\clearpage

\appendix

\setcounter{figure}{0}

\setcounter{table}{0}

\section{Supplementary Information}

\begin{figure}[hp]
  \centering
  \includegraphics[width=\textwidth]{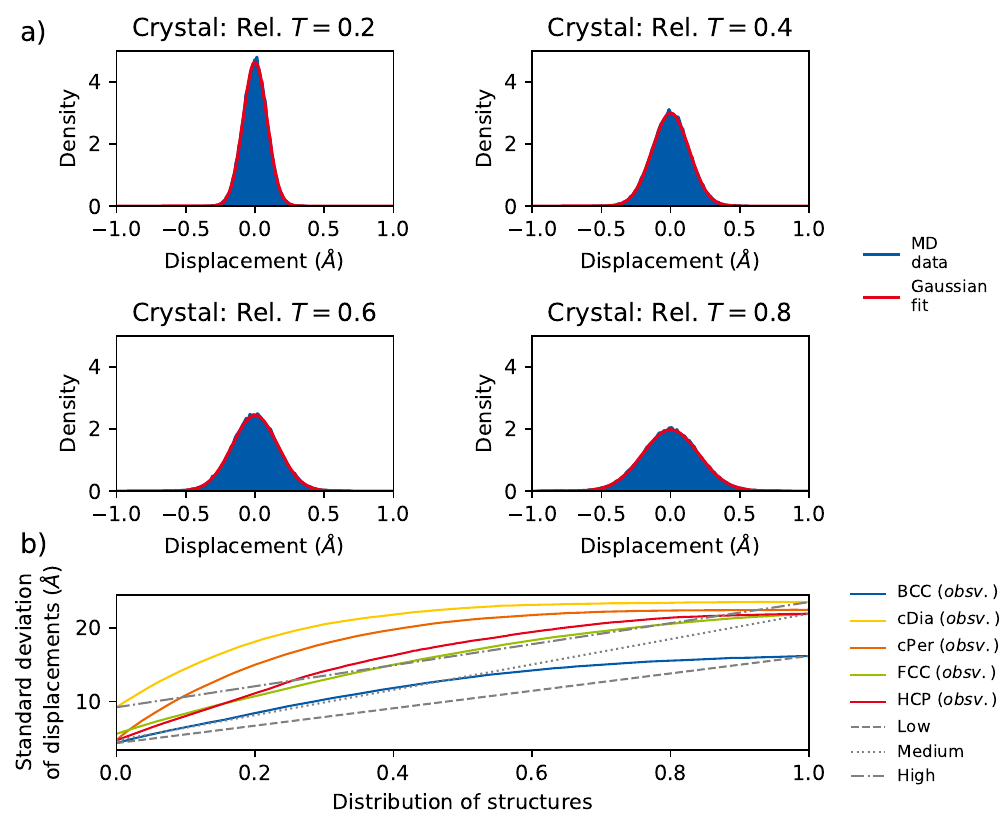}
  \caption{
  (a) Distribution of displacements in the crystalline phase of the cDia sample calculated for the training data MD trajectories. The relative temperature \(\text{Rel.}~T\) corresponding to the temperature regime from \SI{0}{\kelvin} to the homogeneous melting temperature \(T_\text{M,H}\). (b) Standard deviations we assumed for the liquid phase for the cases \textit{low}, \textit{medium} and \textit{high}. In case of \textit{observed} we used an individual displacement amplitude for each structure. 
  \label{figS1}}
\end{figure}

\begin{table}[]
\centering
\caption{Details of the thermodynamic conditions under which the \sio{} training data was generated. We annealed crystalline input structures under NPT conditions from 10~K up to 6000~K with a heating rate of 2$\times$10$^{12}$ K/s.  The pressures of these MD simulations are given in the second column. In the third column, the transitions temperatures are given at which the phase transforms into the other given or an unknown phase (sometimes we observe also amorphisation). The fourth column gives the melting point. Above the melting point snapshots are labeled as molten training data. In the lower part of the table we show the protocol for amorphous training data. We generated sample structures by quenching with different quench rates and under various pressures. The samples have been compressed at constant temperatures (third column) from the starting pressure (first column) to the final pressure (second column).}
\resizebox{0.7\linewidth}{!}{
\begin{tabular}{@{}clll@{}}
\hline
\hline
\multicolumn{1}{l}{Phases}                                                            & Pressure (GPa)       & Transition Temperature (K) & Melting Temperature (K) \\ 
\hline
$\alpha$-cristobalite $\rightarrow$ $\beta$-cristobalite                              & 0                    & 525                        & 4335                    \\
\hline
$\alpha$-moganite $\rightarrow$ $\beta$-moganite                                      & 0                    & 605                        & 4215                    \\
\hline
\multirow{3}{*}{$\alpha$-quartz $\rightarrow$ $\beta$-quartz}                         & 0                    & 885                        & 4275                    \\
                                                                                      & 5                    & 2225                       & 4275                    \\
                                                                                      & 10                   & 3270                       & 4070                    \\
                                                                                      \hline
low tridymite $\rightarrow$ high tridymite                                            & 0                    & 415                        & 4325                    \\
\hline
\multirow{2}{*}{$P3_221$ \cite{badroTheoreticalStudyFivecoordinated1997}}             & 30                   & 945                        & 4880                    \\
                                                                                      & 60                   & 550                        & 5785                    \\
                                                                                      \hline
\multirow{4}{*}{$C2$ \cite{choudhuryInitioStudiesPhonon2006}}                         & 30                   & ---                        & 3695                    \\
                                                                                      & 60                   & 3250                       & ---                     \\
                                                                                      & 90                   & 2735                       & ---                     \\
                                                                                      & 120                  & 2390                       & ---                     \\
                                                                                      \hline
\multirow{3}{*}{d-NiAs-type \cite{liuNewHighpressureModifications1978}}               & 20                   & ---                        & 5230                    \\
                                                                                      & 40                   & ---                        & 5750                    \\
                                                                                      & 60                   & ---                        & ---                     \\
                                                                                      \hline
\multirow{3}{*}{$I2/a$ \cite{tseHighpressureDensificationAmorphous1992}}              & 10                   & ---                        & 3850                    \\
                                                                                      & 40                   & ---                        & 5400                    \\
                                                                                      & 70                   & ---                        & ---                     \\
                                                                                      \hline
\multirow{4}{*}{NaTiF$_4$-type \cite{teterHighPressurePolymorphism1998}}              & 10                   & ---                        & 4200                    \\
                                                                                      & 50                   & ---                        & ---                     \\
                                                                                      & 90                   & ---                        & ---                     \\
                                                                                      & 120                  & ---                        & ---                     \\
                                                                                      \hline
\multirow{4}{*}{$P2_1/c$ \cite{teterHighPressurePolymorphism1998}}                    & 10                   & ---                        & 4175                    \\
                                                                                      & 50                   & ---                        & ---                     \\
                                                                                      & 90                   & ---                        & ---                     \\
                                                                                      & 120                  & ---                        & ---                     \\
                                                                                      \hline
\multirow{3}{*}{$P3_2$ \cite{wentzcovitchNewPhasePressure1998}}                       & 30                   & 1525                       & 4270                    \\
                                                                                      & 60                   & 2180                       & ---                     \\
                                                                                      & 90                   & 2600                       & ---                     \\
                                                                                      \hline
\multirow{2}{*}{$P3_221$ \cite{badroTheoreticalStudyFivecoordinated1997}}             & 30                   & 945                        & 4880                    \\
                                                                                      & 60                   & 550                        & 5785                    \\
                                                                                      \hline
\multirow{2}{*}{$Pnc2$ \cite{svishchevOrthorhombicQuartzlikePolymorph1997}}           & 0                    & 2620                       & 3580                    \\
                                                                                      & 10                   & ---                        & 3825                    \\
                                                                                      \hline
SnO$_2$-type \cite{teterHighPressurePolymorphism1998}                                 & 10                   & ---                        & 4320                    \\
                                                                                      & 50                   & ---                        & ---                     \\
                                                                                      & 90                   & ---                        & ---                     \\
                                                                                      & 120                  & ---                        & ---                     \\
                                                                                      \hline
\multirow{5}{*}{rosiaite-type \cite{otzenEvidenceRosiaitestructuredHighpressure2023}} & 0                    & ---                        & 2645                    \\
                                                                                      & 20                   & 3200                       & 4345                    \\
                                                                                      & 40                   & 3600                       & 5410                    \\
                                                                                      & 60                   & 4100                       & ---                     \\
                                                                                      & 80                   & 4500                       & ---                     \\
                                                                                      \hline
\multirow{3}{*}{CaCl$_2$ $\rightarrow$ stishovite}                                    & 75                   & 650                        & ---                     \\
                                                                                      & 90                   & 1670                       & ---                     \\
                                                                                      & 105                  & 2820                       & ---                     \\
                                                                                      \hline
\multirow{5}{*}{stishovite}                                                           & 0                    & ---                        & 3430                    \\
                                                                                      & 15                   & ---                        & 4920                    \\
                                                                                      & 30                   & ---                        & 5800                    \\
                                                                                      & 45                   & ---                        & ---                     \\
                                                                                      & 60                   & ---                        & ---                     \\
                                                                                      \hline
\multirow{5}{*}{seifertite}                                                            & 0                    & ---                        & 3310                    \\
                                                                                      & 50                   & ---                        & ---                     \\
                                                                                      & 100                  & ---                        & ---                     \\
                                                                                      & 150                  & ---                        & ---                     \\
                                                                                      & 200                  & ---                        & ---                     \\
                                                                                      \hline
\multirow{5}{*}{pyrite}                                                               & 0                    & ---                        & 2090                    \\
                                                                                      & 50                   & ---                        & 4940                    \\
                                                                                      & 100                  & ---                        & ---                     \\
                                                                                      & 150                  & ---                        & ---                     \\
                                                                                      & 200                  & ---                        & ---                     \\
                                                                                      \hline
                                                                                      \hline
\multicolumn{1}{l}{}                                                                  &                      &                            &                         \\
\hline
\hline
\multicolumn{1}{l}{Input structure}                                                   & Start Pressure (GPa) & Final Pressure (GPa)       & Temperatures            \\
\hline

amorphous (quench rate: 10$^{12}$ K/s, p=0 GPa)                                       & 0                    & 200                        & (500,1000,1500,2000)    \\
amorphous (quench rate: 10$^{13}$ K/s, p=0 GPa)                                       & 0                    & 200                        & (500,1000,1500,2000)    \\
amorphous (quench rate: 10$^{12}$ K/s, p=200 GPa)                                     & 200                  & 0                          & (500,1000,1500,2000)    \\
amorphous (quench rate: 10$^{13}$ K/s, p=200 GPa)                                     & 200                  & 0                          & (500,1000,1500,2000)    \\
\hline
\hline
\end{tabular}
}
\label{stab:structures}
\end{table}

\begin{figure}
    \centering
    \includegraphics[width=17cm]{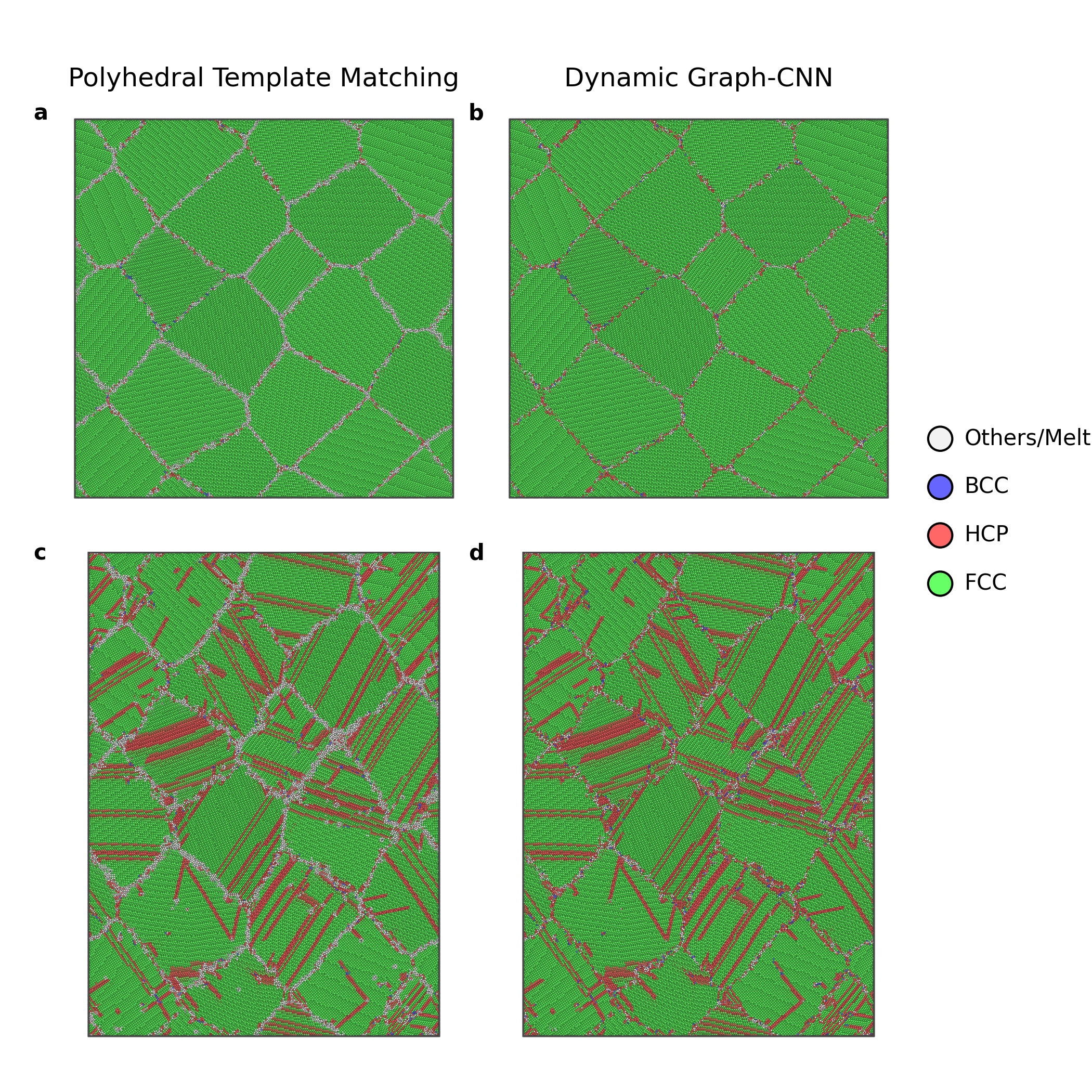}
    \caption{High entropy polycrystal with the composition Ni$_{60}$Mn$_{10}$Co$_{10}$Cr$_{10}$Fe$_{10}$. (a)-(b) shows the polycrystal without load analysed with PTM (RMSD: 0.15) in (a) and the DG-CNN (16 input points) in (b). The grains are classified as FCC by both methods. Moreover, the grain boundaries can be differentiated from grains easily. The PTM predicts the grain boundary atoms mostly to be \textit{Other}, in contrast, most atoms are classified by the DG-CNN as HCP. (c)-(d) shows the polycrystal deformed under tension with stacking faults. Both methods are clearly showing the stacking faults in good agreement. The structures are taken from Ref.~\citenum{nag2024}.}
    \label{fig:hea_polycrystal}
\end{figure}

\begin{figure}[hp]
  \centering
  \includegraphics[width=17cm]{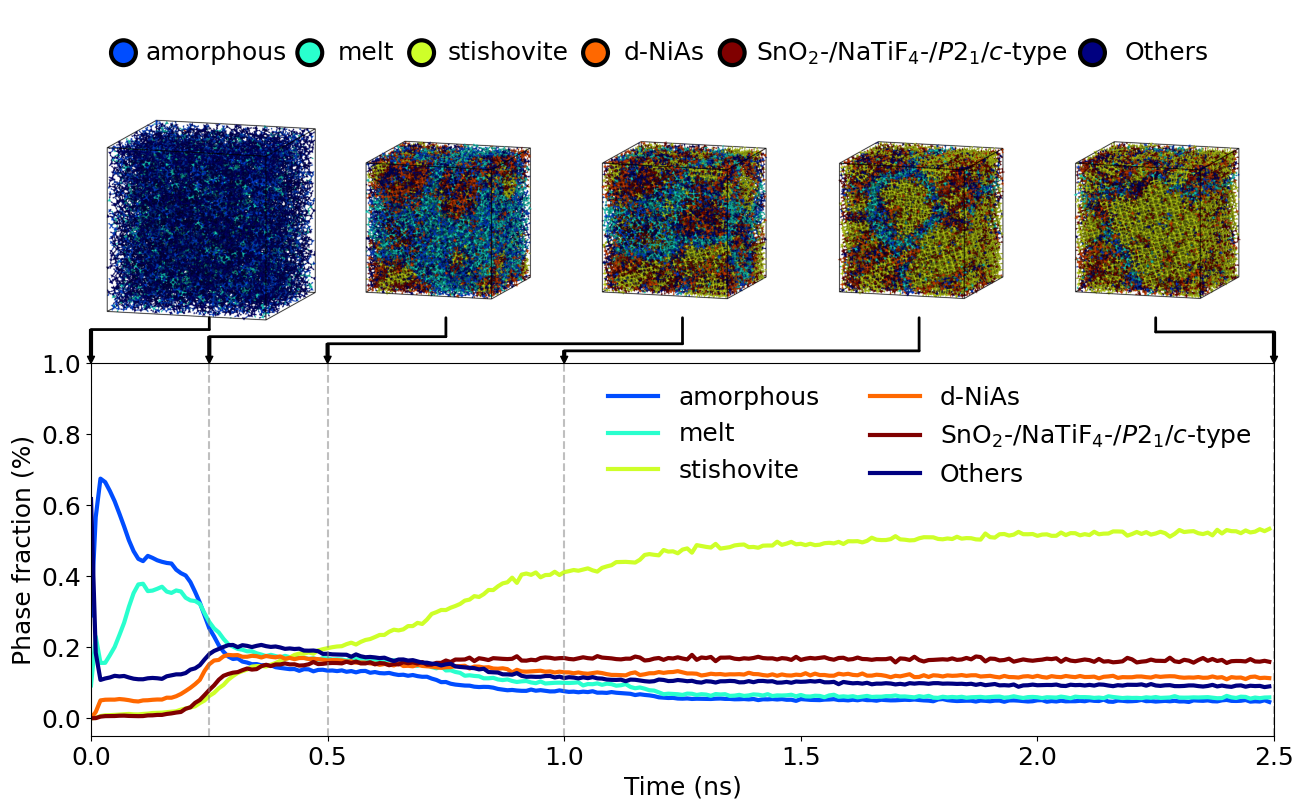}
  \caption{%
  \textbf{a)} Structural evolution in an initially amorphous \sio{} sample which is subjected to shock loading. The DG-CNN recognised structure of each atom is marked by color coding. We used 16 input points. The structure class of SnO$_2$-, NaTiF$_4$- and $P2_1/c$-type silica is summarized as one. All other structure types are summarized at others. During the shock the amorphous sample is strongly compressed, which comes along with an increase in temperature. Moreover, we observe a clear change in the structural classification. (b) Shows the phase fractions during this transformation. First, we observed a strong increase of molten phase, followed by an increase of the d-NiAs phase. Afterwards, the stishovite phase fraction is increasing until it takes up nearly 60\% of the crystal after 2.5~ns simulation time. 
  \label{figS3}}
\end{figure}

\begin{figure}[hp]
  \centering
  \includegraphics[width=17cm]{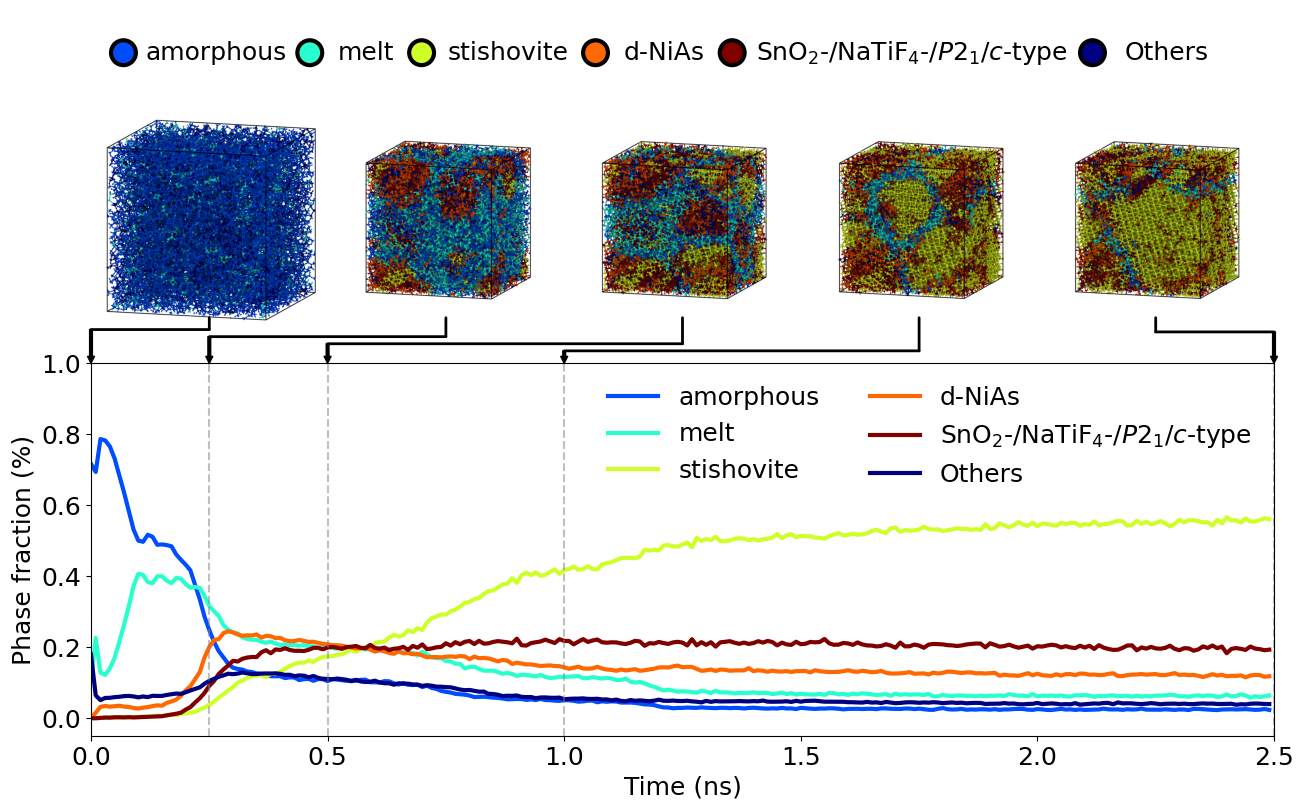}
  \caption{%
  \textbf{a)} Structural evolution in an initially amorphous \sio{} sample which is subjected to shock loading. The DG-CNN recognised structure of each atom is marked by color coding. We used 32 input points. The structure class of SnO$_2$-, NaTiF$_4$- and $P2_1/c$-type silica is summarized as one. All other structure types are summarized at others. During the shock the amorphous sample is strongly compressed, which comes along with an increase in temperature. Moreover, we observe a clear change in the structural classification. (b) Shows the phase fractions during this transformation. First, we observed a strong increase of molten phase, followed by an increase of the d-NiAs phase. Afterwards, the stishovite phase fraction is increasing until it takes up nearly 60\% of the crystal after 2.5~ns simulation time. 
  \label{figS4}}
\end{figure}

\clearpage

\newcommand{\newblock}{}
\bibliographystyle{unsrt}
\bibliography{literature.bib}

\begin{thebibliography}{10}

\bibitem{Honeycutt1987}
J.~Dana. Honeycutt and Hans~C. Andersen.
\newblock Molecular dynamics study of melting and freezing of small
  {{Lennard-Jones}} clusters.
\newblock {\em J. Phys. Chem.}, 91(19):4950--4963, 1987.

\bibitem{Faken1994}
Daniel Faken and Hannes J{\'o}nsson.
\newblock Systematic analysis of local atomic structure combined with {{3D}}
  computer graphics.
\newblock {\em Computational Materials Science}, 2(2):279--286, 1994.

\bibitem{Ackland2006}
G.~J. Ackland and A.~P. Jones.
\newblock Applications of local crystal structure measures in experiment and
  simulation.
\newblock {\em Phys. Rev. B}, 73(5):054104, 2006.

\bibitem{Stukowski2012a}
Alexander Stukowski.
\newblock Structure identification methods for atomistic simulations of
  crystalline materials.
\newblock {\em Modelling Simul. Mater. Sci. Eng.}, 20(4):045021, 2012.

\bibitem{Larsen2016}
Peter~Mahler Larsen, S{\o}ren Schmidt, and Jakob Schi{\o}tz.
\newblock Robust structural identification via polyhedral template matching.
\newblock {\em Modelling Simul. Mater. Sci. Eng.}, 24(5):055007, 2016.

\bibitem{Maras2016}
E.~Maras, O.~Trushin, A.~Stukowski, T.~{Ala-Nissila}, and H.~J{\'o}nsson.
\newblock Global transition path search for dislocation formation in {{Ge}} on
  {{Si}}(001).
\newblock {\em Computer Physics Communications}, 205:13--21, 2016.

\bibitem{Nguyen2015}
Andrew~H. Nguyen and Valeria Molinero.
\newblock Identification of {{Clathrate Hydrates}}, {{Hexagonal Ice}}, {{Cubic
  Ice}}, and {{Liquid Water}} in {{Simulations}}: The {{CHILL}}+ {{Algorithm}}.
\newblock {\em J. Phys. Chem. B}, 119(29):9369--9376, 2015.

\bibitem{Ziletti2018}
Angelo Ziletti, Devinder Kumar, Matthias Scheffler, and Luca~M. Ghiringhelli.
\newblock Insightful classification of crystal structures using deep learning.
\newblock {\em Nat Commun}, 9(1):2775, 2018.

\bibitem{Bartok2010}
Albert~P. Bart{\'o}k, Mike~C. Payne, Risi Kondor, and G{\'a}bor Cs{\'a}nyi.
\newblock Gaussian {{Approximation Potentials}}: {{The Accuracy}} of {{Quantum
  Mechanics}}, without the {{Electrons}}.
\newblock {\em Phys. Rev. Lett.}, 104(13):136403, 2010.

\bibitem{Bartok2013}
Albert~P. Bart{\'o}k, Risi Kondor, and G{\'a}bor Cs{\'a}nyi.
\newblock On representing chemical environments.
\newblock {\em Phys. Rev. B}, 87(18):184115, 2013.

\bibitem{Leitherer2021}
Andreas Leitherer, Angelo Ziletti, and Luca~M. Ghiringhelli.
\newblock Robust recognition and exploratory analysis of crystal structures via
  {{Bayesian}} deep learning.
\newblock {\em Nat Commun}, 12(1):6234, 2021.

\bibitem{Chung2022}
Heejung~W. Chung, Rodrigo Freitas, Gowoon Cheon, and Evan~J. Reed.
\newblock Data-centric framework for crystal structure identification in
  atomistic simulations using machine learning.
\newblock {\em Phys. Rev. Materials}, 6(4):043801, 2022.

\bibitem{Goryaeva2020}
Alexandra~M. Goryaeva, Clovis Lapointe, Chendi Dai, Julien D{\'e}r{\`e}s,
  Jean-Bernard Maillet, and Mihai-Cosmin Marinica.
\newblock Reinforcing materials modelling by encoding the structures of defects
  in crystalline solids into distortion scores.
\newblock {\em Nat Commun}, 11(1):4691, 2020.

\bibitem{beaulieu2024}
Zoé Faure~Beaulieu, Volker~L. Deringer, and Fausto Martelli.
\newblock {High-dimensional order parameters and neural network classifiers
  applied to amorphous ices}.
\newblock {\em The Journal of Chemical Physics}, 160(8):081101, 02 2024.

\bibitem{DeFever2019}
Ryan~S. DeFever, Colin Targonski, Steven~W. Hall, Melissa~C. Smith, and Sapna
  Sarupria.
\newblock A generalized deep learning approach for local structure
  identification in molecular simulations.
\newblock {\em Chem. Sci.}, 10(32):7503--7515, 2019.

\bibitem{Qi2016}
C.~Qi, Hao Su, Kaichun Mo, and Leonidas~J. Guibas.
\newblock {{PointNet}}: {{Deep}} learning on point sets for {{3D}}
  classification and segmentation.
\newblock {\em 2017 IEEE Conf. Comput. Vis. Pattern Recognit. CVPR}, pages
  77--85, 2016.

\bibitem{heaneySilicaPhysicalBehavior1994a}
Peter~J. Heaney, Charles~T. Prewitt, and Gerald~V. Gibbs, editors.
\newblock {\em Silica: {{Physical Behavior}}, {{Geochemistry}}, and {{Materials
  Applications}}}.
\newblock {De Gruyter}, {Berlin, Boston}, December 1994.

\bibitem{badroTheoreticalStudyFivecoordinated1997}
James Badro, David~M. Teter, Robert~T. Downs, Philippe Gillet, Russell~J.
  Hemley, and Jean-Louis Barrat.
\newblock Theoretical study of a five-coordinated silica polymorph.
\newblock {\em Physical Review B}, 56(10):5797--5806, September 1997.

\bibitem{choudhuryInitioStudiesPhonon2006}
N.~Choudhury and S.~L. Chaplot.
\newblock {\emph{Ab Initio}} studies of phonon softening and high-pressure
  phase transitions of {$\alpha$} -quartz {{Si O}} 2.
\newblock {\em Physical Review B}, 73(9):094304, March 2006.

\bibitem{liuNewHighpressureModifications1978}
Lin-Gun Liu, William~A. Bassett, and John Sharry.
\newblock New high-pressure modifications of {{GeO2}} and {{SiO2}}.
\newblock {\em Journal of Geophysical Research: Solid Earth},
  83(B5):2301--2305, 1978.

\bibitem{tseHighpressureDensificationAmorphous1992}
John~S. Tse, Dennis~D. Klug, and Yvon Le~Page.
\newblock High-pressure densification of amorphous silica.
\newblock {\em Physical Review B}, 46(10):5933--5938, September 1992.

\bibitem{teterHighPressurePolymorphism1998}
David~M. Teter, Russell~J. Hemley, Georg Kresse, and J{\"u}rgen Hafner.
\newblock High {{Pressure Polymorphism}} in {{Silica}}.
\newblock {\em Physical Review Letters}, 80(10):2145--2148, March 1998.

\bibitem{wentzcovitchNewPhasePressure1998}
Renata~M. Wentzcovitch, Cesar Da~Silva, James~R. Chelikowsky, and Nadia
  Binggeli.
\newblock A {{New Phase}} and {{Pressure Induced Amorphization}} in {{Silica}}.
\newblock {\em Physical Review Letters}, 80(10):2149--2152, March 1998.

\bibitem{svishchevOrthorhombicQuartzlikePolymorph1997}
Igor~M. Svishchev, Peter~G. Kusalik, and Vladimir~V. Murashov.
\newblock Orthorhombic quartzlike polymorph of silica: {{A}} molecular-dynamics
  simulation study.
\newblock {\em Physical Review B}, 55(2):721--725, January 1997.

\bibitem{otzenEvidenceRosiaitestructuredHighpressure2023}
Christoph Otzen, Hanns-Peter Liermann, and Falko Langenhorst.
\newblock Evidence for a rosiaite-structured high-pressure silica phase and its
  relation to lamellar amorphization in quartz.
\newblock {\em Nature Communications}, 14(1):606, February 2023.

\bibitem{Wang2019}
Yue Wang, Yongbin Sun, Ziwei Liu, Sanjay~E. Sarma, Michael~M. Bronstein, and
  Justin~M. Solomon.
\newblock Dynamic {{Graph CNN}} for {{Learning}} on {{Point Clouds}}.
\newblock {\em ACM Trans. Graph.}, 38(5):1--12, 2019.

\bibitem{Stukowski2010}
Alexander Stukowski.
\newblock Visualization and analysis of atomistic simulation data with
  {{OVITO}}\textendash the {{Open Visualization Tool}}.
\newblock {\em Modelling Simul. Mater. Sci. Eng.}, 18(1):015012, 2010.

\bibitem{Thompson2022}
Aidan~P. Thompson, H.~Metin Aktulga, Richard Berger, Dan~S. Bolintineanu,
  W.~Michael Brown, Paul~S. Crozier, Pieter~J. {in 't Veld}, Axel Kohlmeyer,
  Stan~G. Moore, Trung~Dac Nguyen, Ray Shan, Mark~J. Stevens, Julien Tranchida,
  Christian Trott, and Steven~J. Plimpton.
\newblock {{LAMMPS}} - a flexible simulation tool for particle-based materials
  modeling at the atomic, meso, and continuum scales.
\newblock {\em Computer Physics Communications}, 271:108171, 2022.

\bibitem{Paszke2019}
Adam Paszke, Sam Gross, Francisco Massa, Adam Lerer, James Bradbury, Gregory
  Chanan, Trevor Killeen, Zeming Lin, Natalia Gimelshein, Luca Antiga, Alban
  Desmaison, Andreas Kopf, Edward Yang, Zachary DeVito, Martin Raison, Alykhan
  Tejani, Sasank Chilamkurthy, Benoit Steiner, Lu~Fang, Junjie Bai, and Soumith
  Chintala.
\newblock {{PyTorch}}: {{An}} imperative style, high-performance deep learning
  library.
\newblock In {\em Advances in Neural Information Processing Systems 32}, pages
  8024--8035. {Curran Associates, Inc.}, 2019.

\bibitem{VanDerWalt2011}
St{\'e}fan {van der Walt}, S~Chris Colbert, and Ga{\"e}l Varoquaux.
\newblock The {{NumPy Array}}: {{A Structure}} for {{Efficient Numerical
  Computation}}.
\newblock {\em Comput. Sci. Eng.}, 13(2):22--30, 2011.

\bibitem{Virtanen2020}
Pauli Virtanen, Ralf Gommers, Travis~E. Oliphant, Matt Haberland, Tyler Reddy,
  David Cournapeau, Evgeni Burovski, Pearu Peterson, Warren Weckesser, Jonathan
  Bright, St{\'e}fan~J. {van der Walt}, Matthew Brett, Joshua Wilson, K.~Jarrod
  Millman, Nikolay Mayorov, Andrew R.~J. Nelson, Eric Jones, Robert Kern, Eric
  Larson, C.~J. Carey, {\.I}lhan Polat, Yu~Feng, Eric~W. Moore, Jake
  VanderPlas, Denis Laxalde, Josef Perktold, Robert Cimrman, Ian Henriksen,
  E.~A. Quintero, Charles~R. Harris, Anne~M. Archibald, Ant{\^o}nio~H. Ribeiro,
  Fabian Pedregosa, and Paul {van Mulbregt}.
\newblock {{SciPy}} 1.0: Fundamental algorithms for scientific computing in
  {{Python}}.
\newblock {\em Nat. Methods}, 17(3):261--272, 2020.

\bibitem{Tange2011a}
Ole Tange.
\newblock Gnu parallel 20230622 ('nova kakhovka'), Jun 2023.
\newblock {GNU Parallel is a general parallelizer to run multiple serial
  command line programs in parallel without changing them.}

\bibitem{Mendelev2003}
M.~I. Mendelev, S.~Han, D.~J. Srolovitz, G.~J. Ackland, D.~Y. Sun, and M.~Asta.
\newblock Development of new interatomic potentials appropriate for crystalline
  and liquid iron.
\newblock {\em Philosophical Magazine}, 83(35):3977--3994, 2003.

\bibitem{Stillinger1985}
Frank~H. Stillinger and Thomas~A. Weber.
\newblock Computer simulation of local order in condensed phases of silicon.
\newblock {\em Phys. Rev. B}, 31(8):5262--5271, 1985.

\bibitem{Mendelev2008}
M.I. Mendelev, M.J. Kramer, C.A. Becker, and M.~Asta.
\newblock Analysis of semi-empirical interatomic potentials appropriate for
  simulation of crystalline and liquid {{Al}} and {{Cu}}.
\newblock {\em Philosophical Magazine}, 88(12):1723--1750, 2008.

\bibitem{Sun2006}
D.~Y. Sun, M.~I. Mendelev, C.~A. Becker, K.~Kudin, Tomorr Haxhimali, M.~Asta,
  J.~J. Hoyt, A.~Karma, and D.~J. Srolovitz.
\newblock Crystal-melt interfacial free energies in hcp metals: {{A}} molecular
  dynamics study of {{Mg}}.
\newblock {\em Phys. Rev. B}, 73(2):024116, 2006.

\bibitem{Thomas2005}
BS~Thomas, NA~Marks, and BD~Begg.
\newblock Developing pair potentials for simulating radiation damage in complex
  oxides.
\newblock {\em Nuclear Instruments and Methods in Physics Research Section B:
  Beam Interactions with Materials and Atoms}, 228(1-4):288--292, 2005.

\bibitem{Erhard2023}
Linus~C Erhard, Jochen Rohrer, Karsten Albe, and Volker~L Deringer.
\newblock Modelling atomic and nanoscale structure in the silicon--oxygen
  system through active machine learning.
\newblock {\em Nature Communications}, 15(1):1927, 2024.

\bibitem{kirfelEndingP21Coesite1984a}
A.~Kirfel and G.~Will.
\newblock Ending the ``{{P21}}/a coesite,, discussion.
\newblock {\em Zeitschrift f{\"u}r Kristallographie - Crystalline Materials},
  167(1):287--292, October 1984.

\bibitem{keskarStructuralPropertiesNine1992}
Nitin~R. Keskar and James~R. Chelikowsky.
\newblock Structural properties of nine silica polymorphs.
\newblock {\em Phys. Rev. B}, 46(1):1--13, July 1992.

\bibitem{zhangInsituCrystalStructure2016}
Li~Zhang, Dmitry Popov, Yue Meng, Junyue Wang, Cheng Ji, Bing Li, and Ho-kwang
  Mao.
\newblock In-situ crystal structure determination of seifertite {{SiO2}} at 129
  {{GPa}}: {{Studying}} a minor phase near {{Earth}}'s core{\textendash}mantle
  boundary.
\newblock {\em American Mineralogist}, 101(1):231--234, January 2016.

\bibitem{kuwayamaPyriteTypeHighPressureForm2005}
Yasuhiro Kuwayama, Kei Hirose, Nagayoshi Sata, and Yasuo Ohishi.
\newblock The {{Pyrite-Type High-Pressure Form}} of {{Silica}}.
\newblock {\em Science}, 309(5736):923--925, August 2005.

\bibitem{Mendelev2007}
Mikhail~I. Mendelev, Seungwu Han, Won-joon Son, Graeme~J. Ackland, and David~J.
  Srolovitz.
\newblock Simulation of the interaction between {{Fe}} impurities and point
  defects in {{V}}.
\newblock {\em Phys. Rev. B}, 76(21):214105, 2007.

\bibitem{Murdick2006}
D.~A. Murdick, X.~W. Zhou, H.~N.~G. Wadley, D.~{Nguyen-Manh}, R.~Drautz, and
  D.~G. Pettifor.
\newblock Analytic bond-order potential for the gallium arsenide system.
\newblock {\em Phys. Rev. B}, 73(4):045206, 2006.

\bibitem{Foiles1986}
S.~M. Foiles, M.~I. Baskes, and M.~S. Daw.
\newblock Embedded-atom-method functions for the fcc metals {{Cu}}, {{Ag}},
  {{Au}}, {{Ni}}, {{Pd}}, {{Pt}}, and their alloys.
\newblock {\em Phys. Rev. B}, 33(12):7983--7991, 1986.

\bibitem{Mendelev2007a}
M.~I. Mendelev and G.~J. Ackland.
\newblock Development of an interatomic potential for the simulation of phase
  transformations in zirconium.
\newblock {\em Philosophical Magazine Letters}, 87(5):349--359, 2007.

\bibitem{Ravelo2004}
R.~Ravelo, B.~L. Holian, T.~C. Germann, and P.~S. Lomdahl.
\newblock Constant-stress hugoniostat method for following the dynamical
  evolution of shocked matter.
\newblock {\em Phys. Rev. B}, 70:014103, Jul 2004.

\bibitem{Erhard2022}
Linus~C. Erhard, Jochen Rohrer, Karsten Albe, and Volker~L. Deringer.
\newblock A machine-learned interatomic potential for silica and its relation
  to empirical models.
\newblock {\em npj Computational Materials}, 8(1):90, April 2022.

\bibitem{Kingma2015}
Diederik~P. Kingma and Jimmy Ba.
\newblock Adam: {{A}} method for stochastic optimization.
\newblock In Yoshua Bengio and Yann LeCun, editors, {\em 3rd {{Int}}. {{Conf}}.
  {{Learn}}. {{Represent}}. {{ICLR}} 2015 {{San Diego CA USA May}} 7-9 2015
  {{Conf}}. {{Track Proc}}.}, 2015.

\bibitem{Loshchilov2017}
Ilya Loshchilov and Frank Hutter.
\newblock {{SGDR}}: {{Stochastic}} gradient descent with warm restarts.
\newblock In {\em 5th {{Int}}. {{Conf}}. {{Learn}}. {{Represent}}. {{ICLR}}
  2017 {{Toulon Fr}}. {{April}} 24-26 2017 {{Conf}}. {{Track Proc}}.}
  {OpenReview.net}, 2017.

\bibitem{Erhart2013}
Paul Erhart, Jaime Marian, and Babak Sadigh.
\newblock Thermodynamic and mechanical properties of copper precipitates in
  {$\alpha$} -iron from atomistic simulations.
\newblock {\em Phys. Rev. B}, 88(2):024116, 2013.

\bibitem{nag2024}
Shankha Nag, Sriram Anand, and Karsten Albe.
\newblock Deformation of nanocrystalline fcc complex concentrated alloys.
\newblock {\em in preparation}, 2024.

\bibitem{prydeSequencePhaseTransitions1998}
A.~K.~A. Pryde and M.~T. Dove.
\newblock On the {{Sequence}} of {{Phase Transitions}} in {{Tridymite}}.
\newblock {\em Physics and Chemistry of Minerals}, 26(2):171--179, December
  1998.

\bibitem{ramanAvTransformationQuartz1940}
C.~V. Raman and T.~M.~K. Nedungadi.
\newblock The {$\alpha$}-{$\beta$} {{Transformation}} of {{Quartz}}.
\newblock {\em Nature}, 145(3665):147--147, January 1940.

\bibitem{leadbetterTransitionCristobalitePhases1976}
A.~J. Leadbetter and T.~W. Smith.
\newblock The {$\alpha$}{\textemdash}{$\beta$} transition in the cristobalite
  phases of {{SiO2}} and {{AIPO4 II}}. {{Calorimetric}} studies.
\newblock {\em The Philosophical Magazine: A Journal of Theoretical
  Experimental and Applied Physics}, 33(1):113--119, January 1976.

\bibitem{Tracy2018}
Sally~June Tracy, Stefan~J. Turneaure, and Thomas~S. Duffy.
\newblock In situ {{X-Ray Diffraction}} of {{Shock-Compressed Fused Silica}}.
\newblock {\em Physical Review Letters}, 120(13):135702, March 2018.

\bibitem{Shen2016}
Yuan Shen, Shai~B Jester, Tingting Qi, and Evan~J Reed.
\newblock Nanosecond homogeneous nucleation and crystal growth in
  shock-compressed sio2.
\newblock {\em Nature materials}, 15(1):60--65, 2016.

\bibitem{gleasonUltrafastVisualizationCrystallization2015}
A.~E. Gleason, C.~A. Bolme, H.~J. Lee, B.~Nagler, E.~Galtier, D.~Milathianaki,
  J.~Hawreliak, R.~G. Kraus, J.~H. Eggert, D.~E. Fratanduono, G.~W. Collins,
  R.~Sandberg, W.~Yang, and W.~L. Mao.
\newblock Ultrafast visualization of crystallization and grain growth in
  shock-compressed {{SiO2}}.
\newblock {\em Nature Communications}, 6(1):8191, September 2015.

\bibitem{Fecht1992}
H.~J. Fecht.
\newblock Defect-induced melting and solid-state amorphization.
\newblock {\em Nature}, 356(6365):133--135, 1992.

\bibitem{Akiba2019}
Takuya Akiba, Shotaro Sano, Toshihiko Yanase, Takeru Ohta, and Masanori Koyama.
\newblock Optuna: {{A Next-generation Hyperparameter Optimization Framework}},
  2019.

\end{thebibliography}

\end{document}